\documentclass[dvips]{acta} 
\usepackage{supertabular,lscape,epsfig} 
\usepackage{amssymb} 
\usepackage{amsmath}

\SetPages{0}{0}

\SetVol{1}{1}

\begin{document}

\begin{Titlepage}

\Title{Period-Luminosity Relations for Ellipsoidal Binary Stars in the OGLE-III Fields of the Large
Magellanic Cloud\footnote{Based on observations obtained with the 1.3-m
Warsaw telescope at the Las Campanas Observatory of the Carnegie
Institution for Science.}}

\Author{M.~~P~a~w~l~a~k$^1$,\hspace{2pt}
I.~~S~o~s~z~y~\'{n}~s~k~i$^1$,\hspace{2pt}
P.~~P~i~e~t~r~u~k~o~w~i~c~z$^1$,\hspace{2pt}
J.~~S~k~o~w~r~o~n$^1$,\hspace{2pt}
R.~~P~o~l~e~s~k~i$^{1,2}$,\hspace{2pt}
A.~~U~d~a~l~s~k~i$^1$,\hspace{2pt} 
M.~K.~~S~z~y~m~a~\'{n}~s~k~i$^1$,\hspace{2pt} 
M.~~K~u~b~i~a~k$^1$,\hspace{2pt}
G.~~P~i~e~t~r~z~y~\'{n}~s~k~i$^{1,3}$,\hspace{2pt}
\L.~~W~y~r~z~y~k~o~w~s~k~i$^{1,4}$,\hspace{2pt}
K.~~U~l~a~c~z~y~k$^1$,\hspace{2pt} 
S.~~K~o~z~\l~o~w~s~k~i$^1$,\hspace{2pt}
and\hspace{2pt} D.~M.~~S~k~o~w~r~o~n$^1$}
{$^1$Warsaw University Observatory, Al. Ujazdowskie 4, 00-478 Warszawa, Poland\\
e-mail: (mpawlak, soszynsk, pietruk)@astrouw.edu.pl\\
$^2$Department of Astronomy, The Ohio State University, 140 West 18th Avenue, Columbus, OH 43210, USA\\
$^3$Departamento de Astronom{\'i}a, Universidad de Concepci\'{o}n, Casilla 160-C, Chile\\
$^4$Institute of Astronomy, University of Cambridge, Madingley Road, Cambridge~CB3~0HA,~UK
}

\Received{Month Day, Year}
\end{Titlepage}

%
%
\Abstract{We report the discovery of two distinct types of ellipsoidal binary systems occupying, so
called, sequence~E on the period-luminosity (P-L) diagram. We propose that steeper
P-L relation is composed of giant-dwarf binaries, while the other consists of giant-giant binary
systems.
Analysis is based on a sample of 5334 objects, which we select from the OGLE-III survey data towardthe the Large Magellanic Cloud.

We show that one of the components of ellipsoidal binaries is typically either a Red Clump or a Red Giant Branch star, which leads to 
clear separation split of the sequence~E at P = 40 d. In its short-period part, we identify two subsequences corresponding to the
two types of binary systems (E$_2$ and E$_3$), while in the longer-period part the two groups merge forming a single subsequence E$_1$. 

We extract a group of 271 ellipsoidal systems with eccentric orbits, from our sample. We
present the period-luminosity relation they follow.
}
{Stars: binaries: ellipsoidal, Stars: variables:general}

\section{Introduction} 
Period-luminosity (P-L) relations for variable stars represent an important 
tool for astrophysical studies as they allow to determine the absolute
magnitudes of stars based on their periods.
This is a key to distance determination
to the studied objects (\eg Caputo \etal 2002, Dall'Ora \etal 2004, Del Principe \etal 2006, Riess \etal 2012),
but also provides information on their evolutionary status (\eg Riebel \etal 2010) and physical parameters (\eg Fiorentino \etal 2007).

The best known period-luminosity relations are those for pulsating stars: Classical Cepheids 
(\eg Gieren \etal 1998, Udalski \etal 1999, Fouqu{\'e} \etal 2007),
RR Lyrae stars (\eg Bono \etal 2001, Soszy{\'n}ski \etal 2003, Catelan \etal 2004), 
Long Period Variables (\eg Wood \etal 1999, Soszy{\'n}ski \etal 2007).
However, close binary systems are also expected to show this sort of relations,
as the orbital periods are correlated with sizes of the stars, and therefore with 
their luminosities. Ruci{\'n}ski (1994, 2004), Ruci{\'n}ski and Maceroni (2011), and Eker \etal (2009) studied 
period-color-luminosity relations for short-period eclipsing binaries of the W~UMa type.
The obtained relation, expressing the absolute magnitude as a linear function of logarithm of orbital period and 
dereddened color, allows obtaining the absolute magnitude value with an accuracy of about 0.25 mag.
Correlation between the luminosity and period for low-mass X-ray binaries was described
by van Paradijs and McClintock (1994) and Revnivtsev \etal (2012). Optical and infrared absolute
magnitudes of such systems can be described as a function of the X-ray luminosity and orbital period.

Close binary systems, in which one of the components is a red giant deformed due to tidal forces, form a period-luminosity 
relation. This relation, known as sequence E, was discovered by Wood \etal (1999) for stars in the Large Magellanic Cloud (LMC). 
The orbital periods range from a few days up to 1000 days.
Ellipsoidal shape of a star results in changes of the observed brightness of 
the binary at different orbital phases. Light curves of such objects are typically close to sinusoidal, with two 
equal maxima and two minima of different depth per orbital period. 
Period-luminosity relations of such stars was analyzed by Soszy{\'n}ski \etal(2004, 2007),
Fraser \etal (2005), Derekas \etal (2006) and Muraveva \etal (2014). 

Sequence E has its continuation, known as sequence D, toward longer periods.
It is formed by enigmatic Long Secondary Period (LSP) variables.
The LSP phenomenon is observed in about 25\% to 50\% pulsating Asymptotic Giant Branch stars. 
The continuation was first noticed by Soszy{\'n}ski \etal (2004). 
The relation between sequences E and D was also studied by Wood \etal (2004), 
Derekas \etal (2006), Soszy{\'n}ski (2007), and Nicholls \etal (2010, 2011). Soszy{\'n}ski and Udalski (2014) show
that the binary scenario is a possible explanation of the LSP effect. 
A detailed analysis of selected sequence E
ellipsoidal binaries with eccentric orbits was performed by Nicholls and Wood (2012).
Recently, Kim \etal (2014) published a large set of binary stars containing candidates for ellipsoidal systems.

In this work, we perform an independent search for stars lying on sequence E in the data collected in the LMC
during the third phase of the long-term sky variability survey, the Optical Gravitational Lensing Experiment (OGLE-III, Udalski 2003).
We identify 5334 likely candidates for ellipsoidal binaries including 1565 high-confidence identifications. 
We analyze the period-luminosity relations they follow and suggest a physical explanation to the observed sequences.

The structure of the paper is as follows. In Section~2, we describe
the observations and data reductions. Section~3 
describes the process of selection and classification of the ellipsoidal
variables. Section~4 presents the analysis of the sample.
In Section~5, we present 
ellipsoidal systems with eccentric orbits.
Finally, Section~6 summarizes our results.

\section{Observations and Data Reduction}

The photometric data presented in this work were obtained 
with the 1.3-m Warsaw telescope at Las Campanas Observatory, Chile. 
The observatory is operated by the Carnegie Institution for Science.
The telescope was equipped with an eight-chip CCD camera, with the field of view of $35\arcm\times35\arcm$. 
More technical details can be found in Udalski (2003).

The observations lasted for eight years, from June 2001 to May 2009, and were taken
in filters $I$ (90\%) and $V$ (10\%). The covered area in the LMC was about 40 square degrees
and contained approximately $3.6\times10^7$ sources with brightness $12 < I < 21$~mag.
Typically, about 500 epochs per object in the $I$ band were secured.
Photometric measurements were obtained with the Differential Image Analysis (DIA)
technique (Alard and Lupton 1998, Alard 2000, Wo\'{z}niak 2000). 
The data reduction is described in detail in Udalski \etal (2008).

\section{Selection and Classification}

All stars in the OGLE-III fields of the LMC were checked for periodic variability using a
discrete Fourier transform algorithm.
The {\sc Fnpeaks}\footnote{{\it http://helas.astro.uni.wroc.pl/deliverables.php?lang=en\&active=fnpeaks}} code by Z. Ko{\l}aczkowski 
was used.
Objects with the signal-to-noise ($S/N$) of the best period larger than 5.0 were selected for further analysis. 
The $S/N$ parameter obtained with {\sc Fnpeaks} is defined as a ratio of the maximum value in the power
spectrum periodogram to its average value.

Using $I$- and $V$-band mean magnitudes we calculated the Wesenheit index, $W_I$,
given with the formula:
\begin{equation}
W_I = I - 1.55 (V - I) 
\end{equation}
assuming the mean ratio of total to selective absorption $A_I/E(V-I) = 1.55$.
We constructed a ${\log}P$-$W_I$ diagram, which we subsequently used to select objects 
for visual inspection. We analyzed the light curves of stars located in the period-luminosity stripe
corresponding to the sequence E studied by Wood \etal (1999)
and Soszy{\'n}ski \etal (2004, 2007) and its continuation toward shorter periods down to 1~d.
Objects found during previous variability searches were added to the sample.

A total of 19 594 light curves were subjected to visual inspection. We selected
5334 light curves consistent with ellipsoidal variability (the full sample) out of which we 
choose a subset of 1565 high-confidence objects, i.e., the ones with close to sinusoidal light curves
with one minimum deeper than the other one, which is typical for ellipsoidal binaries.
We excluded obvious eclipsing binaries since they would affect the statistical
distribution of both the mean magnitudes and amplitudes. However, some of our ellipsoidal binaries may
have shallow eclipses, which is hard do distinguish from ellipsoidal variability.
A few examples of high-confidence ellipsoidal variables are presented in Fig.~1.

\begin{figure}[p]
\includegraphics[angle=270,width=62mm]{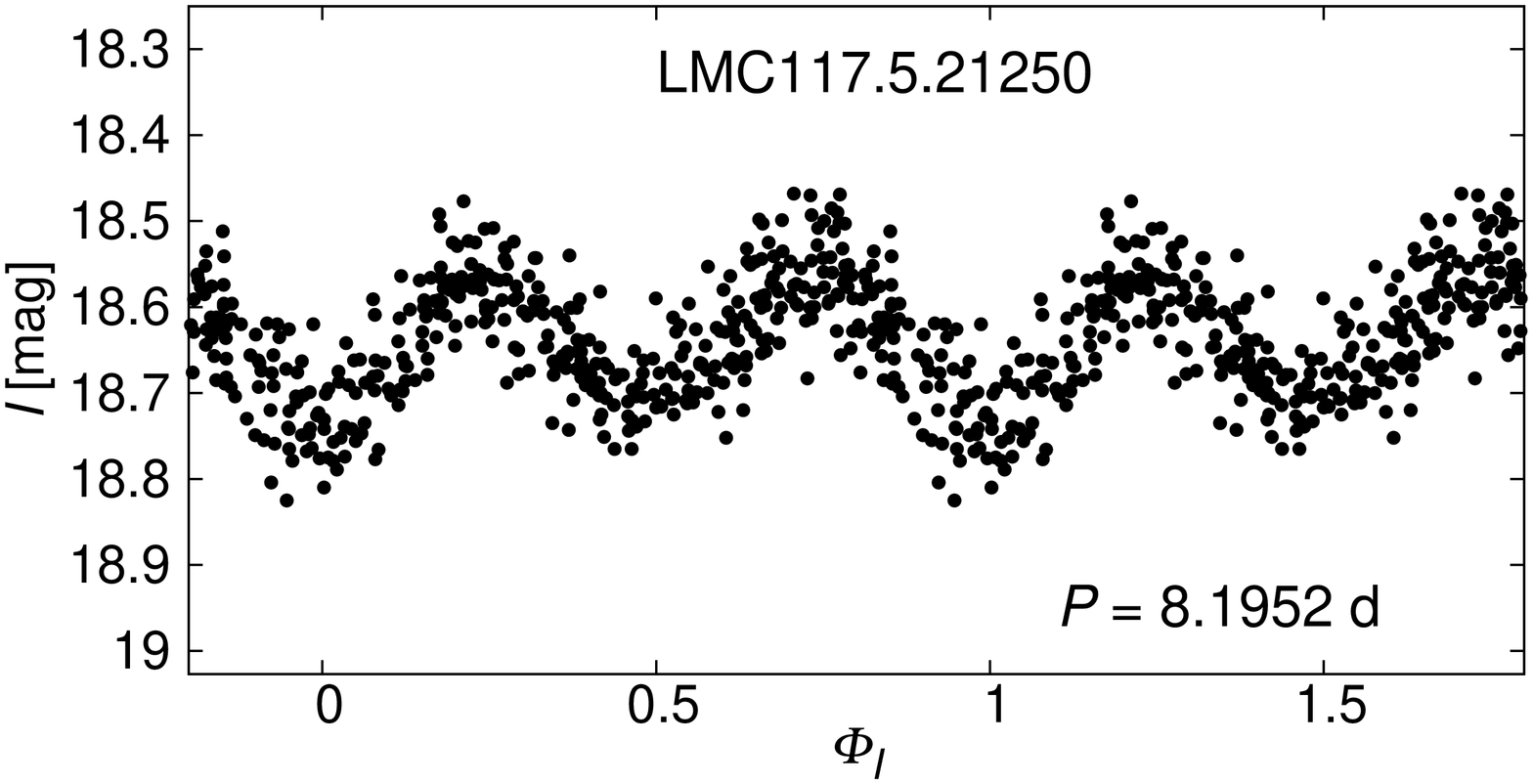}\hfill \includegraphics[angle=270,width=62mm]{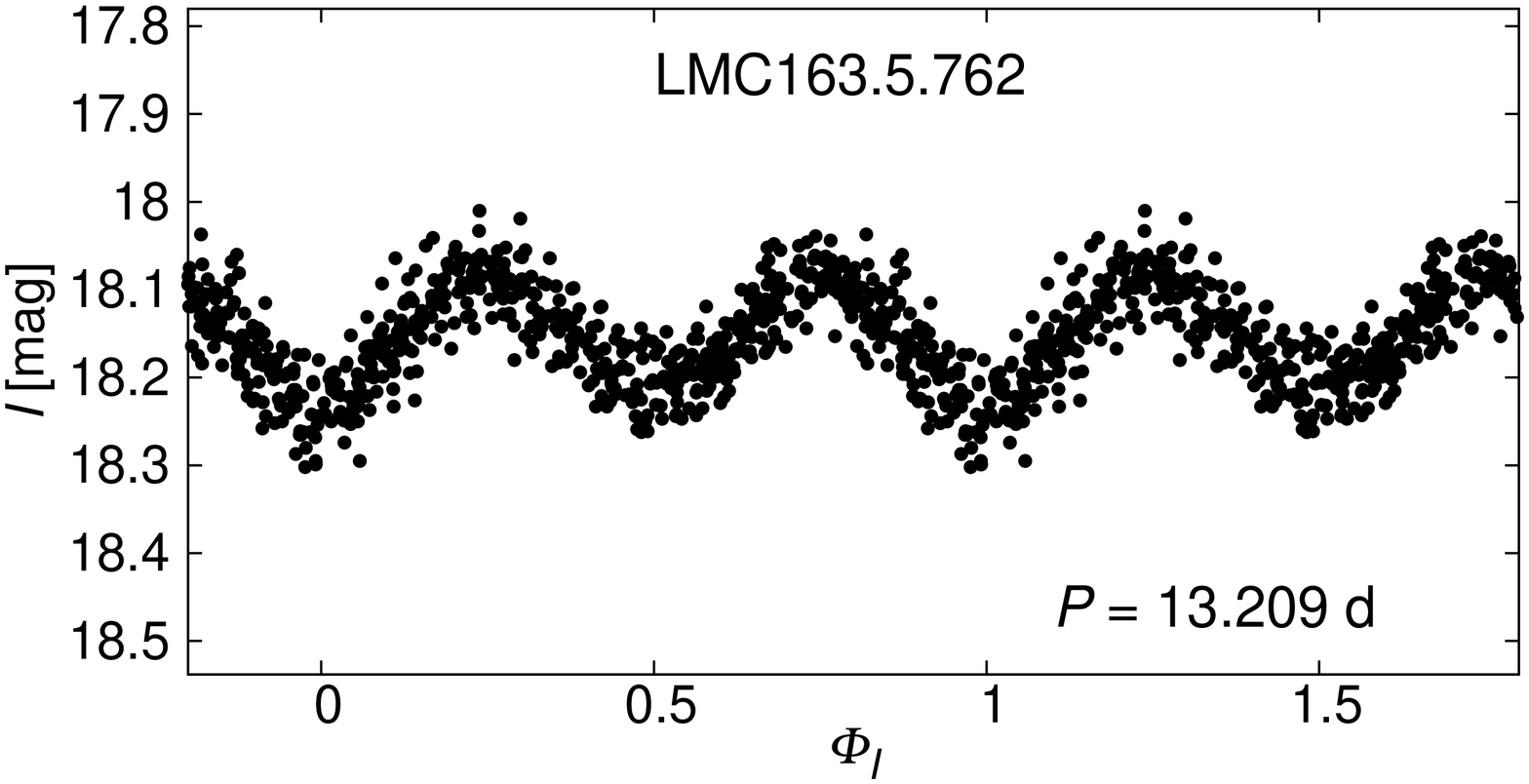} \\
\includegraphics[angle=270,width=62mm]{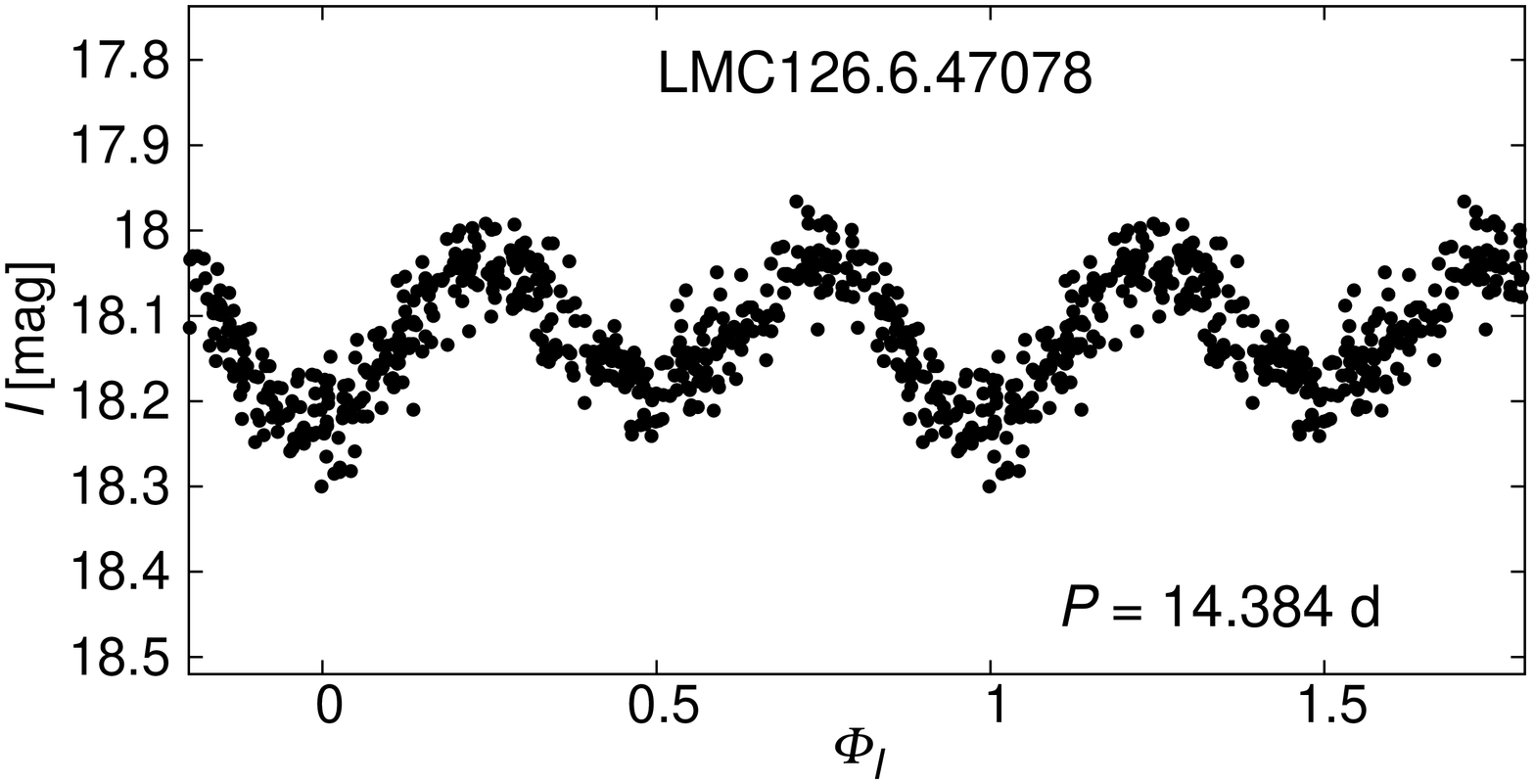}\hfill \includegraphics[angle=270,width=62mm]{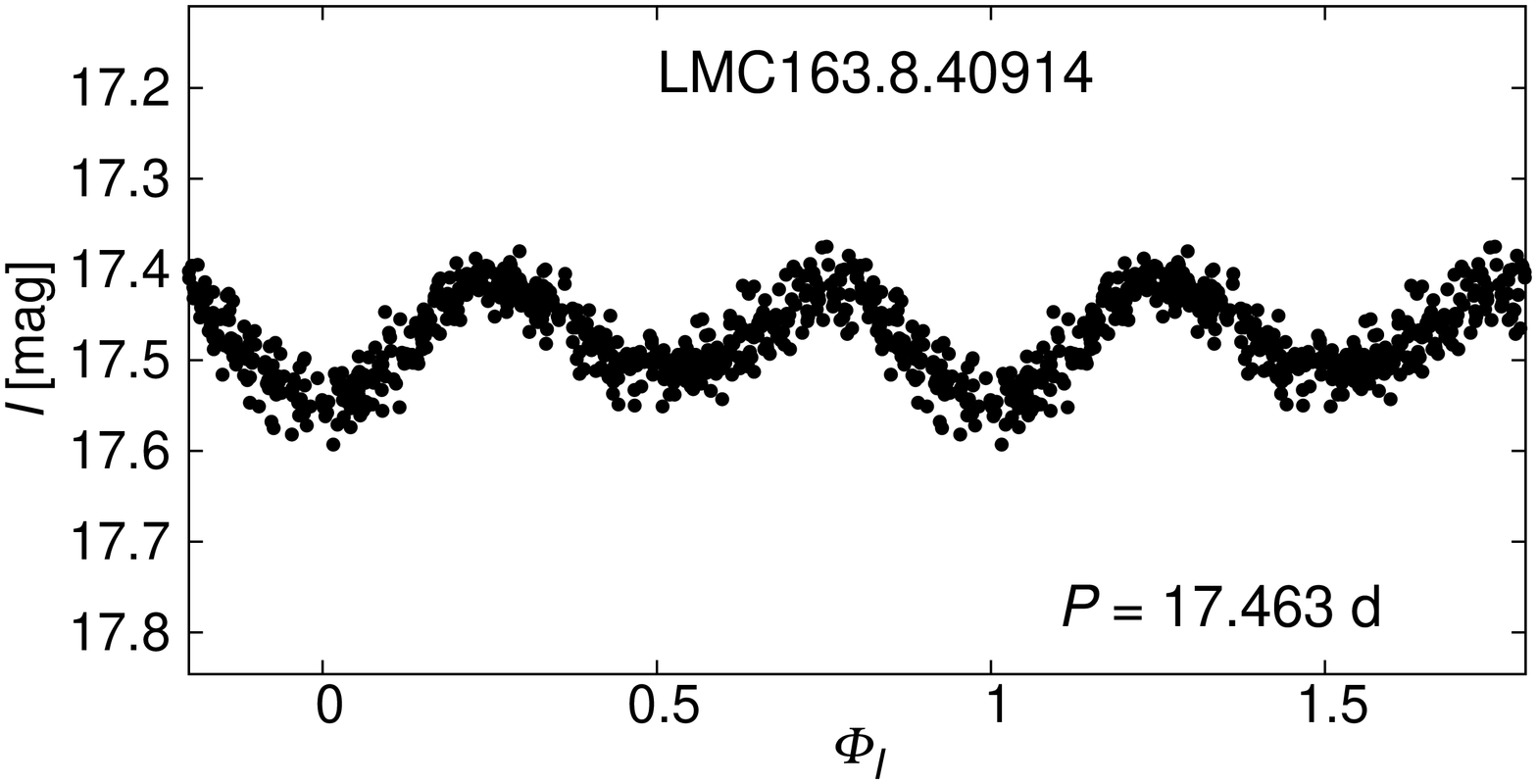} \\
\includegraphics[angle=270,width=62mm]{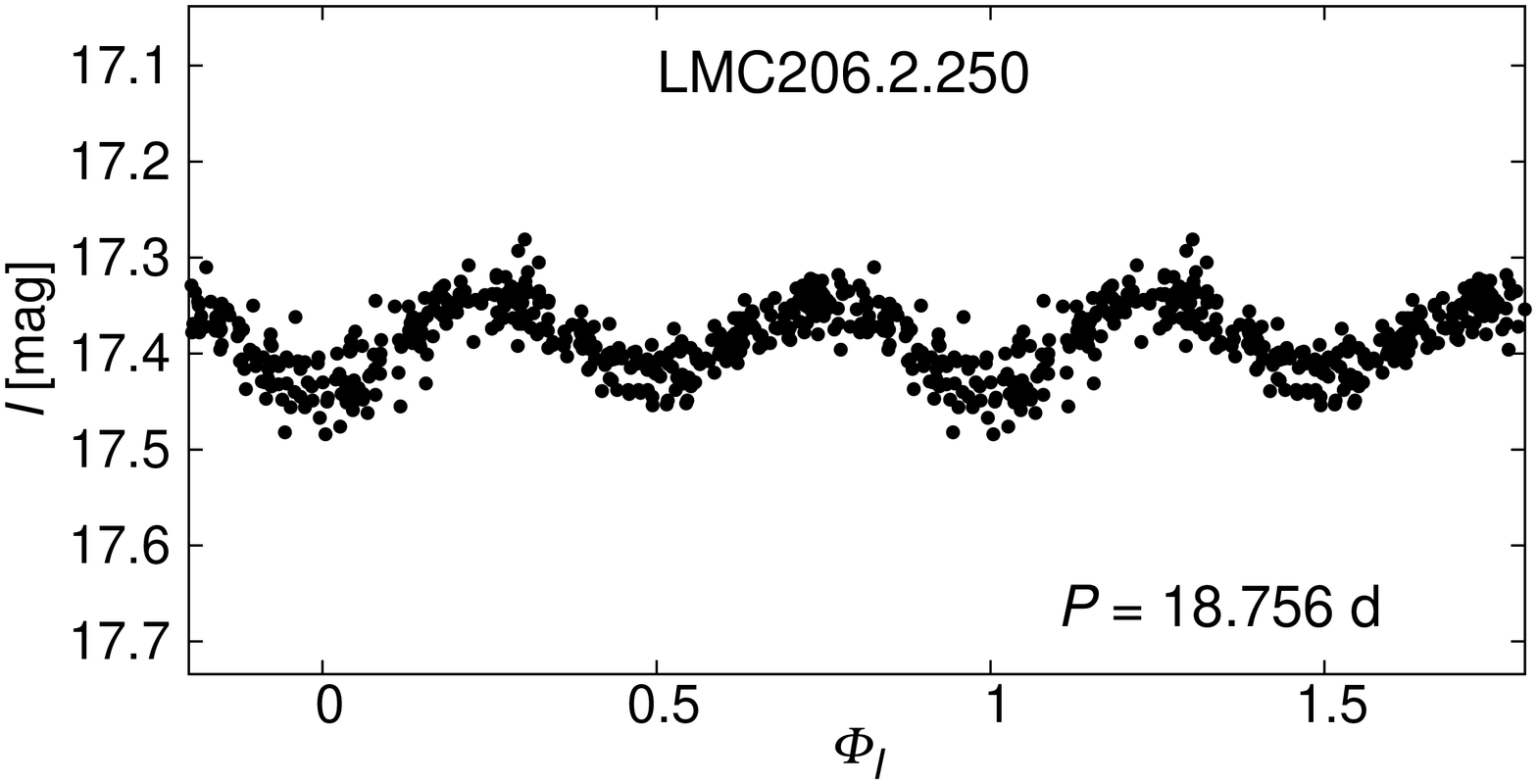}\hfill \includegraphics[angle=270,width=62mm]{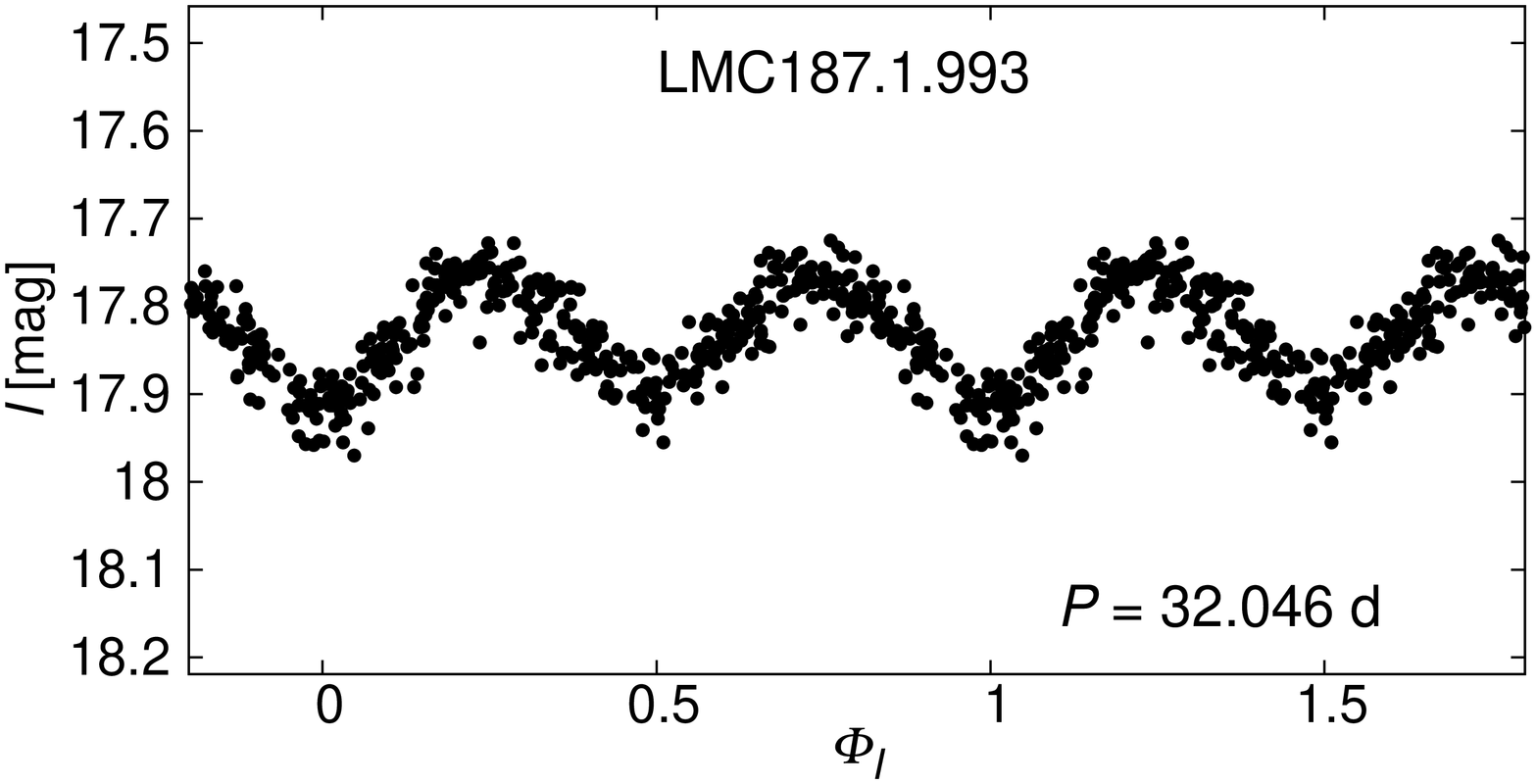} \\
\includegraphics[angle=270,width=62mm]{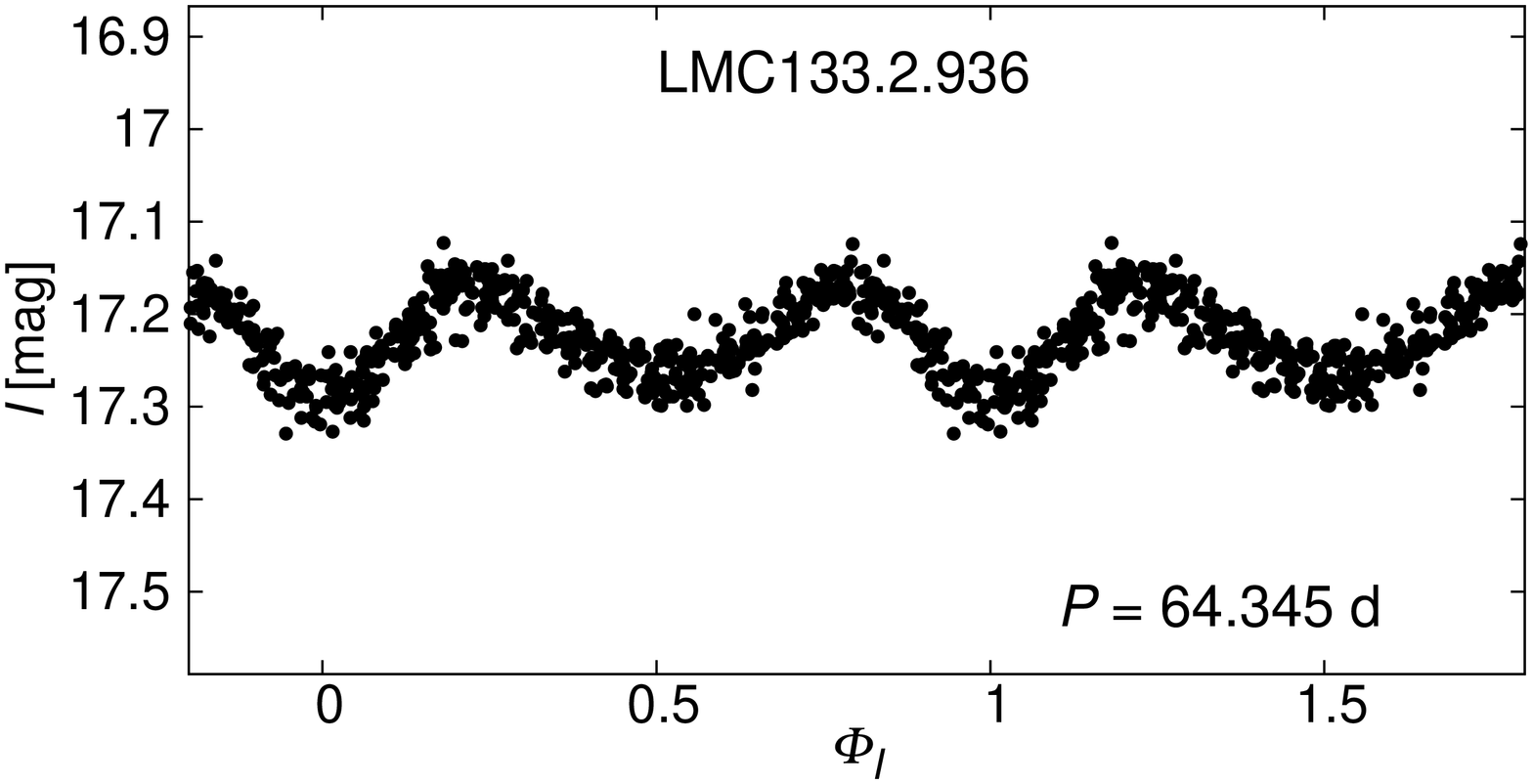}\hfill \includegraphics[angle=270,width=62mm]{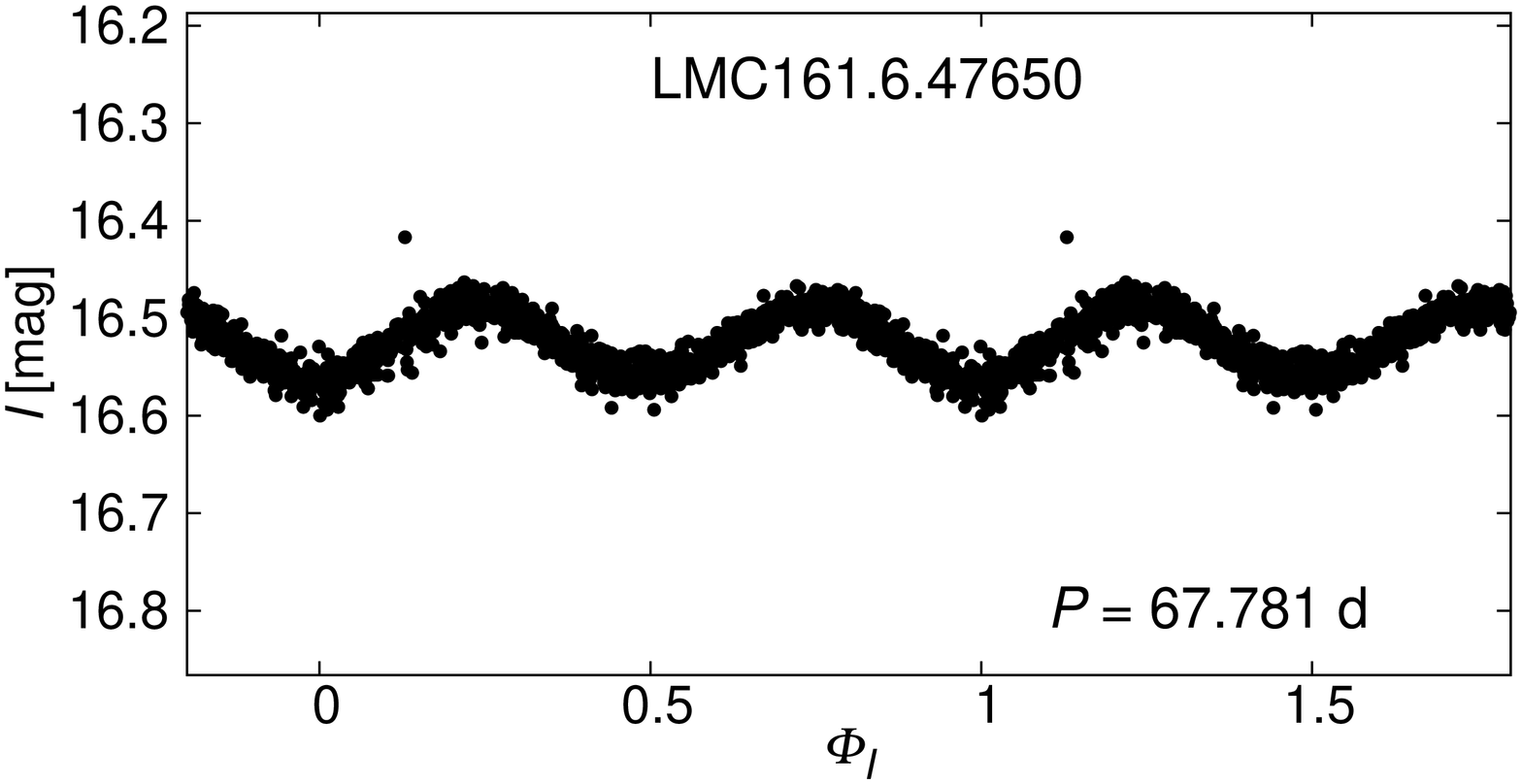} \\
\includegraphics[angle=270,width=62mm]{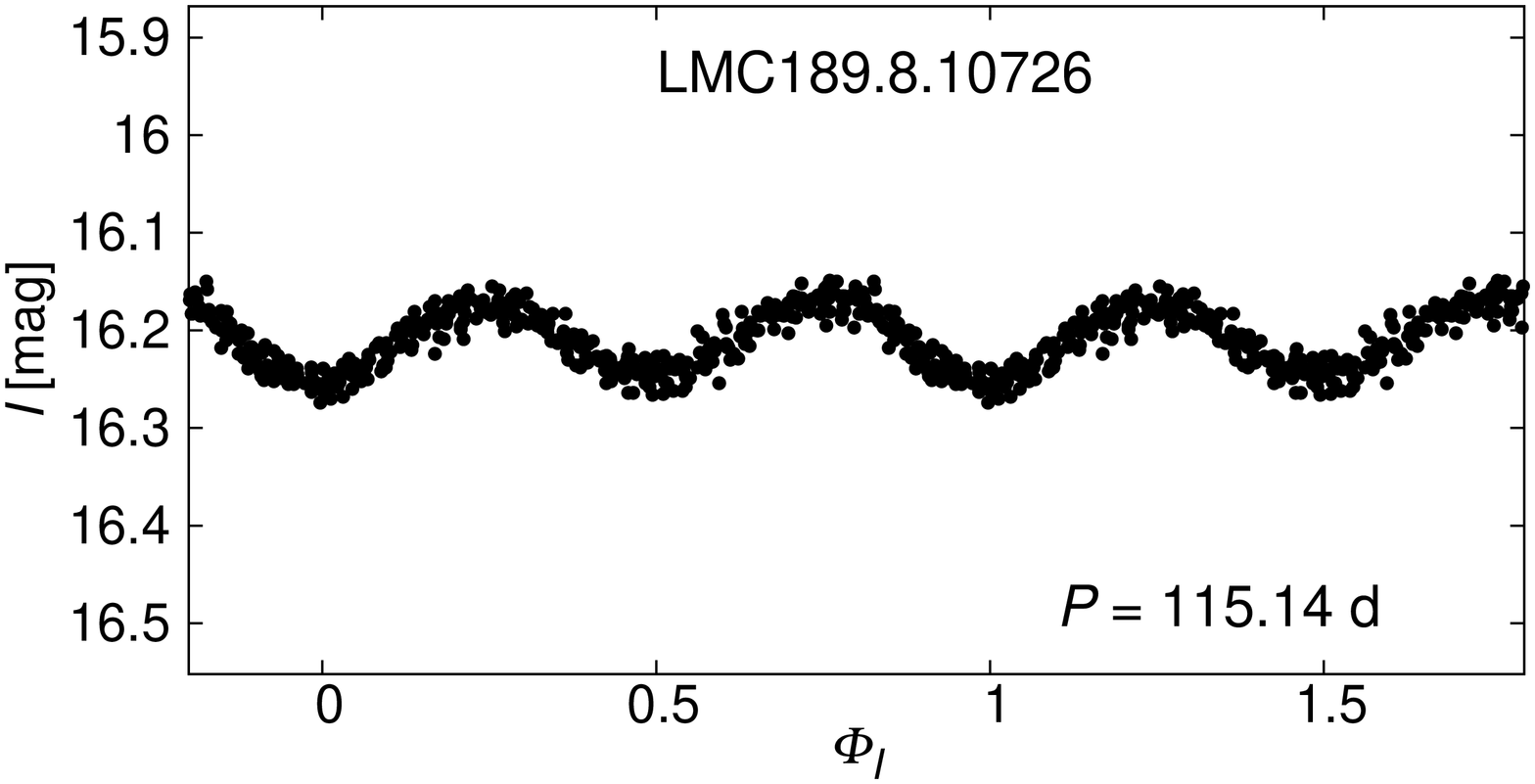}\hfill \includegraphics[angle=270,width=62mm]{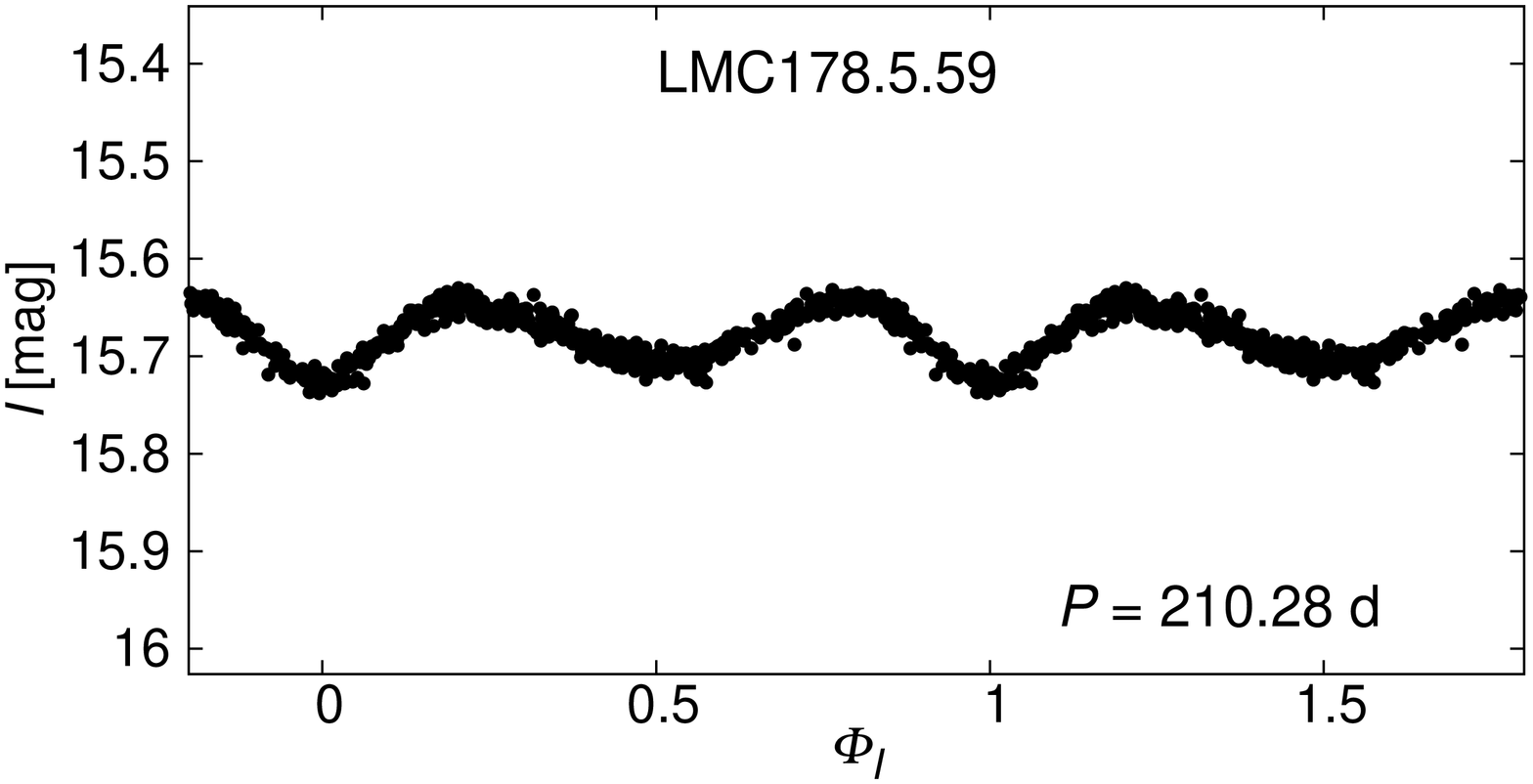} 
\FigCap{Examples of phased light curves of ellipsoidal variables in the LMC, with periods in ascending order.
Systems with longer periods have on average lower amplitudes.}
\end{figure}

\section{Analysis of the Ellipsoidal Variables in the LMC}  

The selected ellipsoidal variables have orbital periods ranging 
from $P=1.3$~d to 1156~d, with the majority between 10~d and 400~d.
Distribution of the orbital periods is shown in Fig.~2. It is clear
that stars group in two ranges of periods: the first with the maximum around 12~d and 
the second around 130~d. Between them, at 
about 40~d, the number of objects is significantly lower. 
The same division can be found in the $\log P - W_I$ diagram presented in Fig.~3.

\begin{figure}[htb]
\centerline{\includegraphics[angle=270,width=130mm]{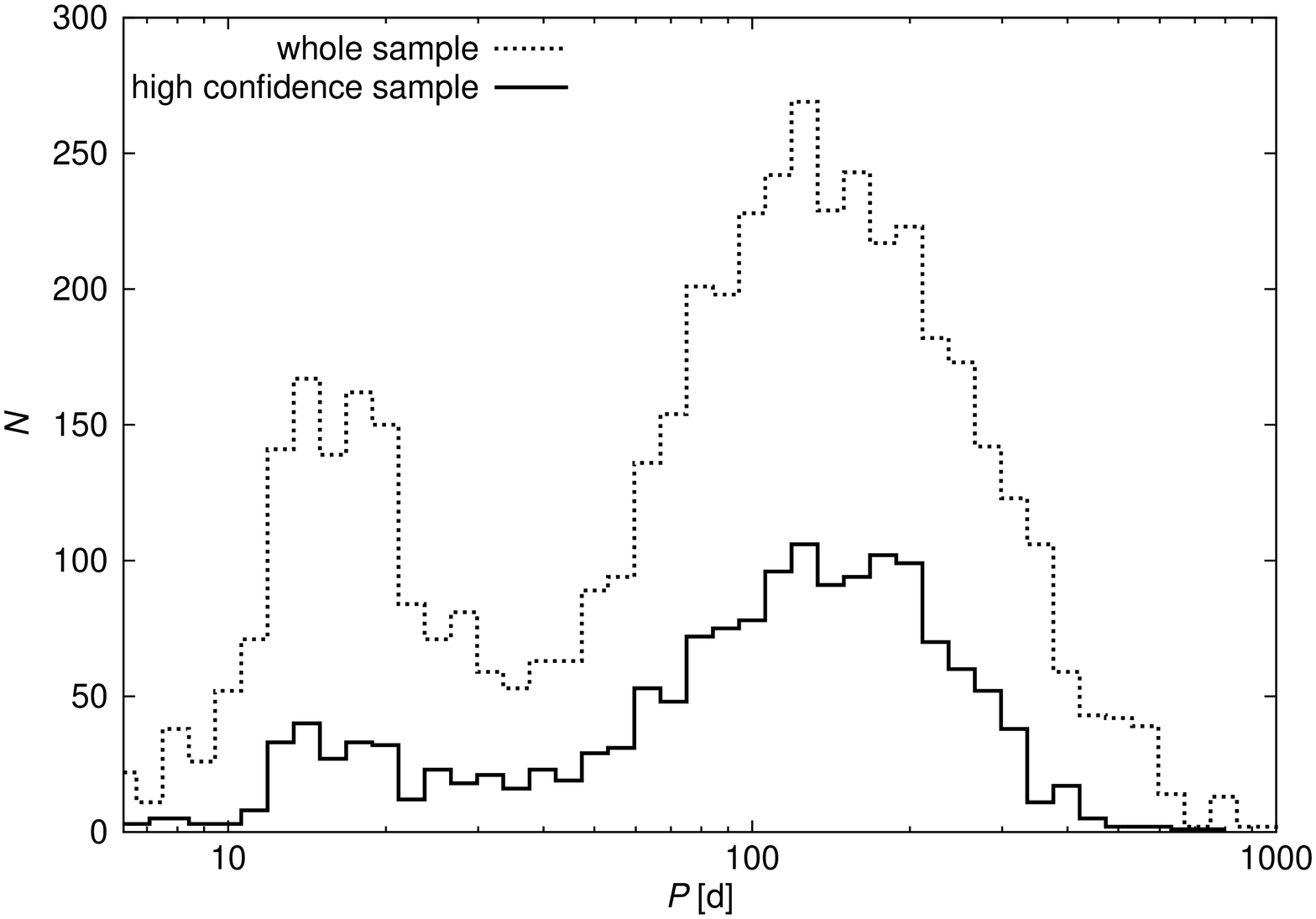}}
\FigCap{Distribution of periods for high-confidence sample of ellipsoidal binaries (solid line) and for the whole sample (dotted line).}
\end{figure}

\begin{figure}[htb]
\centerline{\includegraphics[angle=270,width=130mm]{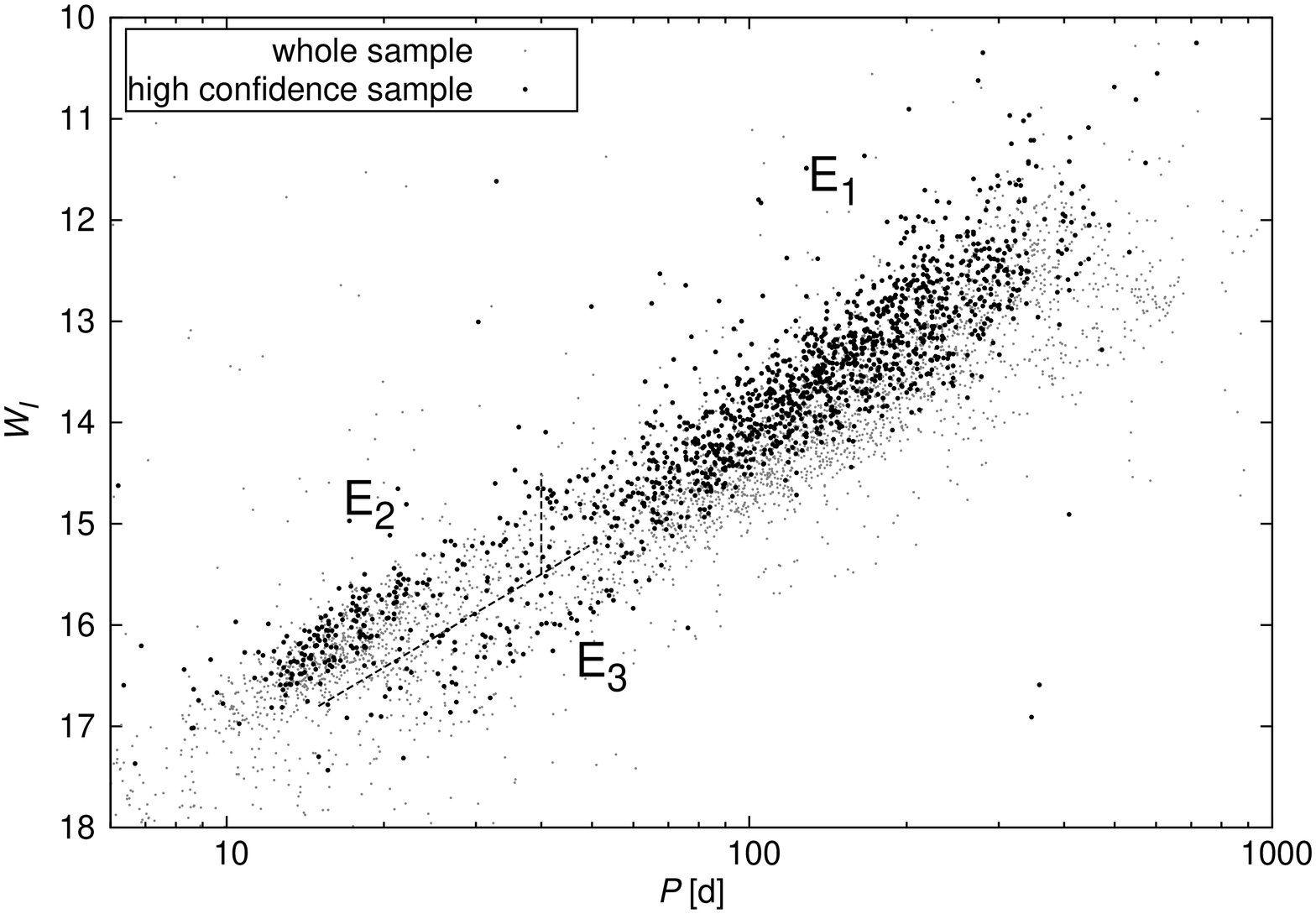}}
\FigCap{Period-luminosity diagram for high-confidence sample of ellipsoidal binaries (large black points) and for the whole sample (small gray points).
Division into three subsequences E$_1$, E$_2$, and E$_3$ is shown.}
\end{figure}

The distribution of stars in the color-magnitude diagram (CMD) 
indicates further difference between the two groups (Fig.~4). Objects with periods longer than
40~d occupy mostly the Red Giant Branch (RGB), while the ones with shorter 
periods lie mostly in the Red Clump (RC) region of the diagram. There is also
a small group of systems located on the Main Sequence (MS) and between 
MS and RGB.

The fraction of ellipsoidal variables relative to the total number of observed stars
in the region of the RC and RGB is shown in Fig.~5. Again, there are two maxima in the distribution,
one around $W_I = 16.0$ mag, which corresponds to the upper region of RC and the second one at about 13.5~mag.
There is also a depletion in the number of stars between the two groups at about 15~mag. 
This distribution shows that there is an
excess of ellipsoidal variables both in the RC and RGB regions of the CMD.

\begin{figure}[htb]
\centerline{\includegraphics[angle=270,width=130mm]{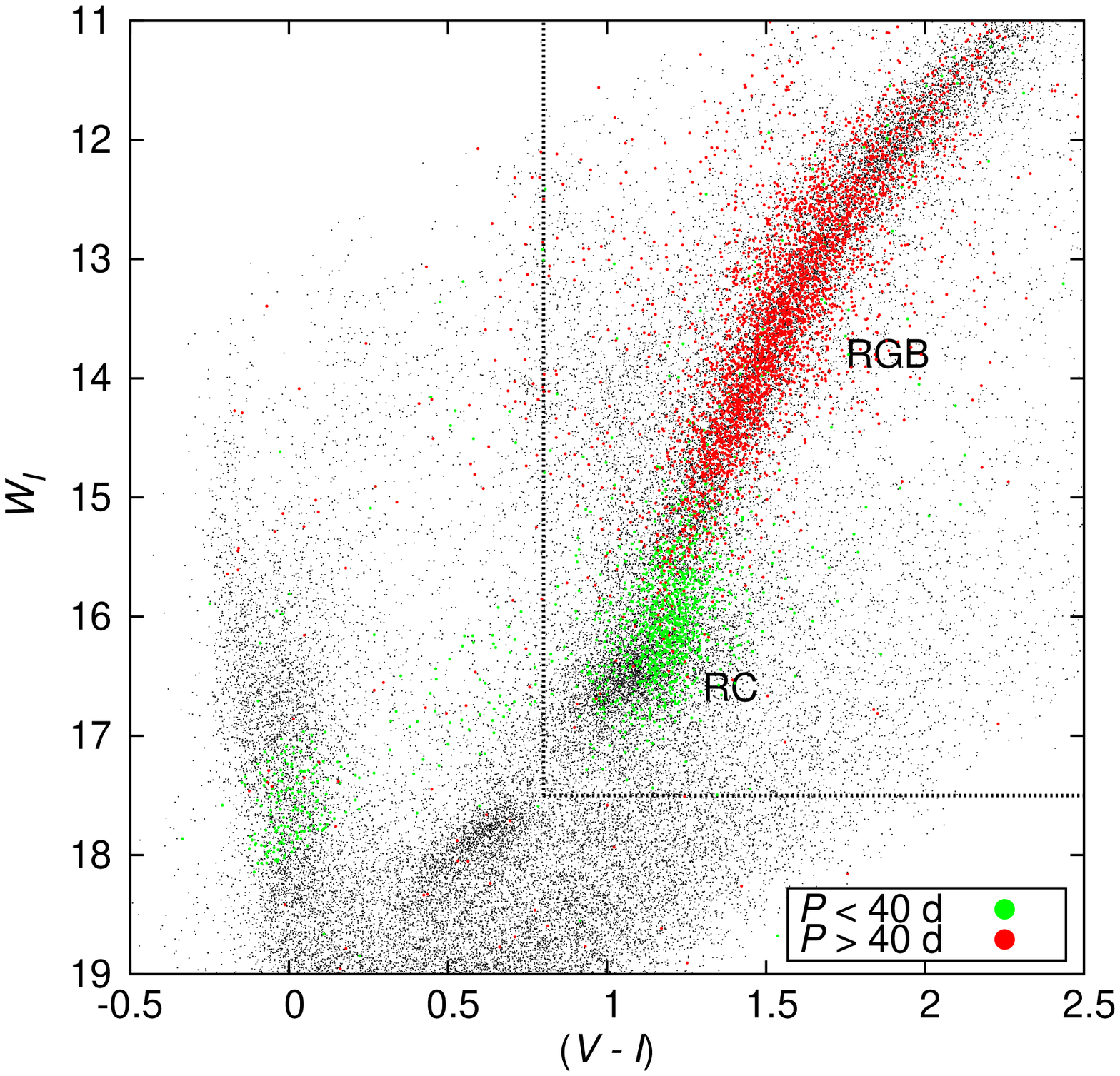}}
\FigCap{Color-magnitude diagram for candidate ellipsoidal variables. Stars with $P < 40$~d are marked in green
and with $P > 40$~d in red.  Field stars come from the photometric
maps of the LMC (Udalski \etal 2008b). The RC and RGB region is marked with the dashed line.}
\end{figure}

\begin{figure}[htb]
\centerline{\includegraphics[angle=270,width=130mm]{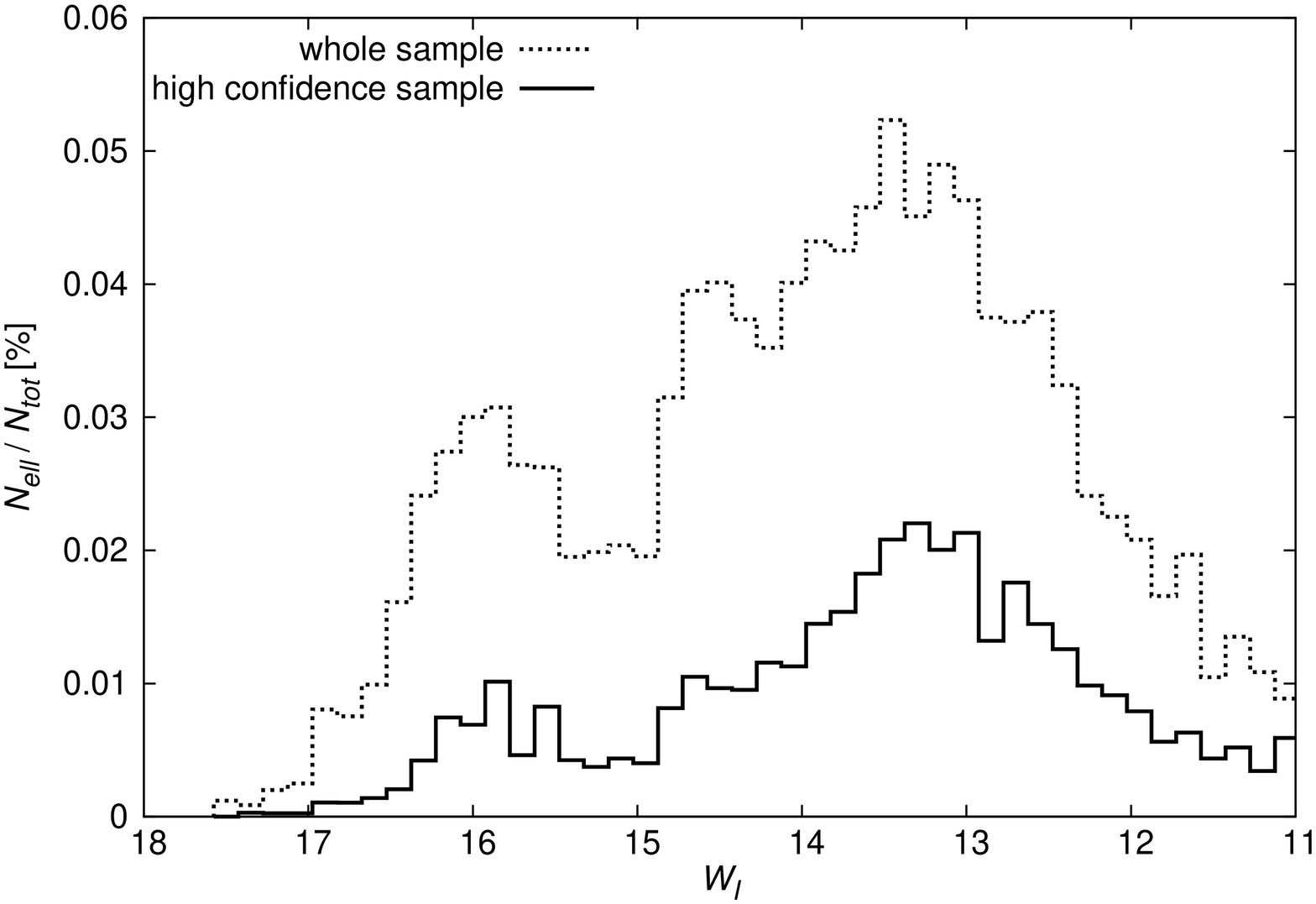}}
\FigCap{Fraction of ellipsoidal variable candidates to all stars in the region of RC and RGB (as marked on Fig.~4) as a function of $W_I$ -- high-confidence (solid line), whole sample (dotted line).}
\end{figure}

The RGB and RC stars form period-luminosity relations shown in Fig.~3. The one formed 
by the RGB stars corresponds to the original sequence E (Wood \etal 1999) for red giants. The relation for 
RC ellipsoidal binaries has not been studied before. We denote relations for the RGB and RC stars as E$_1$ and E$_2$, respectively.

Apart from these two groups, there is another relation, E$_3$, in the period range from 
15~d to 60~d, with $W_I$ fainter by $~1$ mag than E$_2$.
It is not very prominent for the whole sample, but becomes clearly visible while looking at
high-confidence sample of ellipsoidal variables. Shapes of the light curves of these stars are similar to other ellipsoidal
binaries based on both the visual inspection and Fourier parametrization comparison.

\begin{figure}[htb]
\centerline{\includegraphics[angle=270,width=130mm]{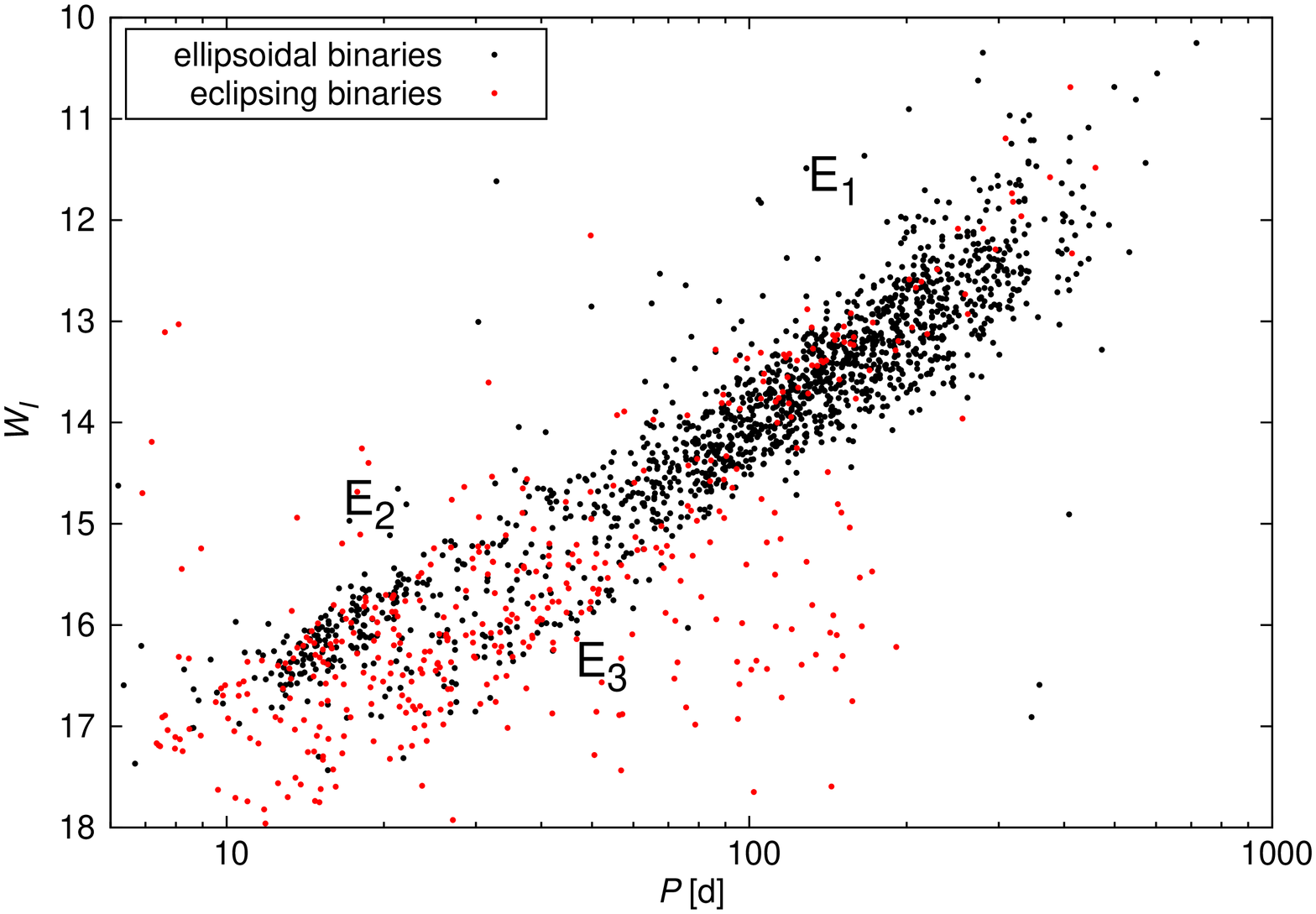}}
\FigCap{Comparison between ellipsoidal (black points) and close eclipsing binaries (red points). While subsequences E$_1$ and E$_2$ are dominated by
ellipsoidal binaries without eclipses, E$_3$ is mostly formed by eclipsing ones.}
\end{figure}
 
In Fig.~6, we present the comparison between our sample of high-confidence ellipsoidal 
binaries and close eclipsing systems identified by Graczyk \etal (2011). Eclipsing binaries also form similar 
sequences. However, we notice that subsequence E$_3$ can be seen even better in eclipsing binaries. 

\begin{figure}[htb]
\centerline{\includegraphics[angle=270,width=130mm]{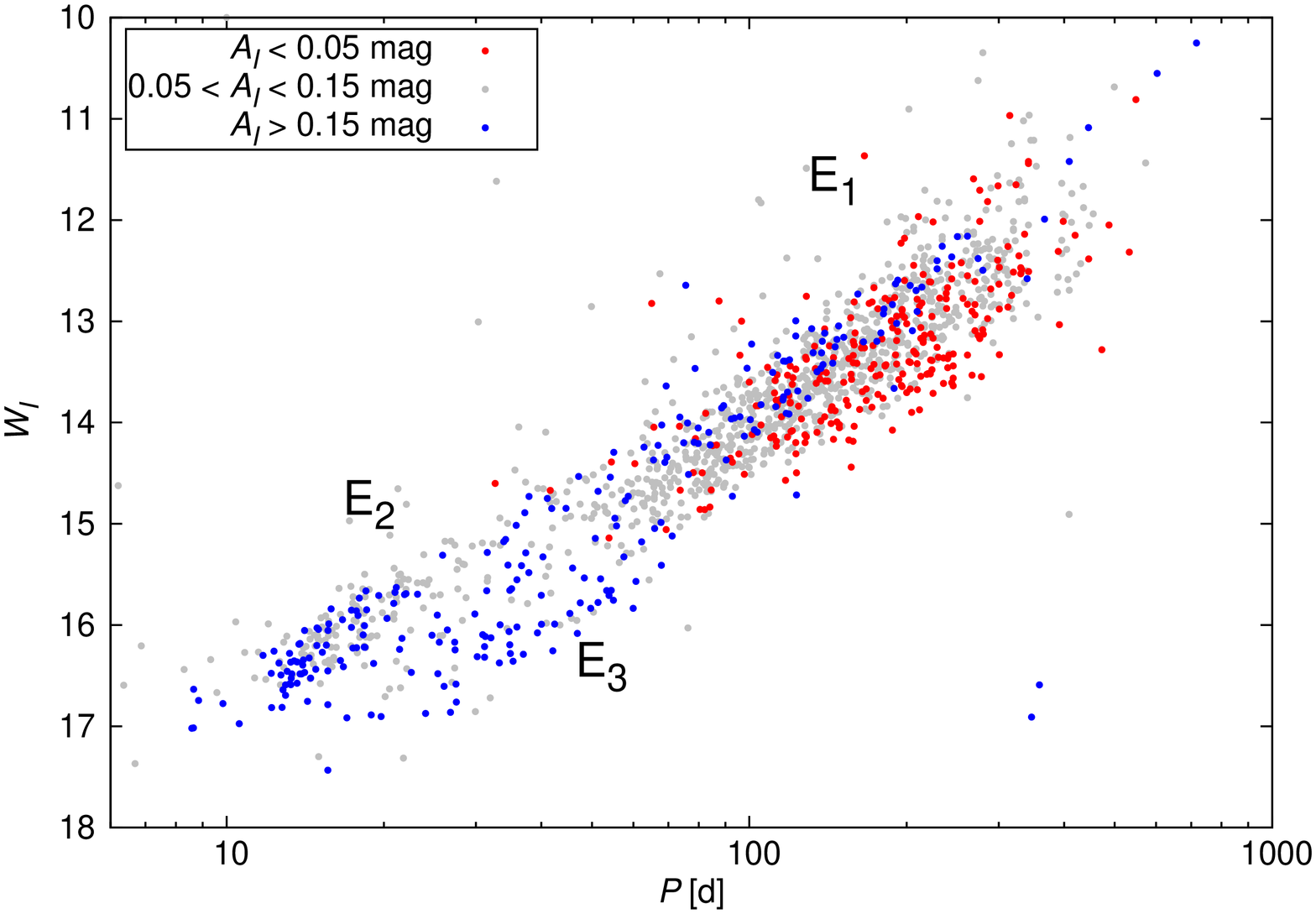}}
\FigCap{Amplitudes in the $I$-band of stars lying in different parts of the period-luminosity diagram. Blue, gray, and red points
represents high-, intermediate-, and low-amplitude objects respectively.}
\end{figure}

Full $I$-band amplitudes of stars located in different parts of the period-luminosity diagram 
are shown in Fig.~7.
Stars from subsequence E$_1$ have on average the lowest amplitudes.
Most of them have intermediate amplitudes ($0.05 < A_I < 0.15$~mag), while there is also a group 
of systems with very low amplitude ($A_I < 0.05$~mag), which are not observed in subsequences E$_2$ and E$_3$.
However, there are also stars in E$_1$ with amplitudes
over 0.15 mag. These stars follow a tighter period-luminosity relation, which was already
noted by Soszy{\'n}ski \etal (2004).
Almost all stars in subsequence E$_3$ have high amplitudes ($A_I > 0.15$~mag), 
while subsequence E$_2$ is composed of both intermediate and high-amplitude stars.
However, it should be noted the completeness of the sample may be lower for the faintest objects. As shown in Fig.~8, 
systems with $W_I = 17$~mag and $A_I<0.05$~mag would be hard to detect with the available photometry.  

\begin{figure}[htb]
\centerline{\includegraphics[angle=270,width=130mm]{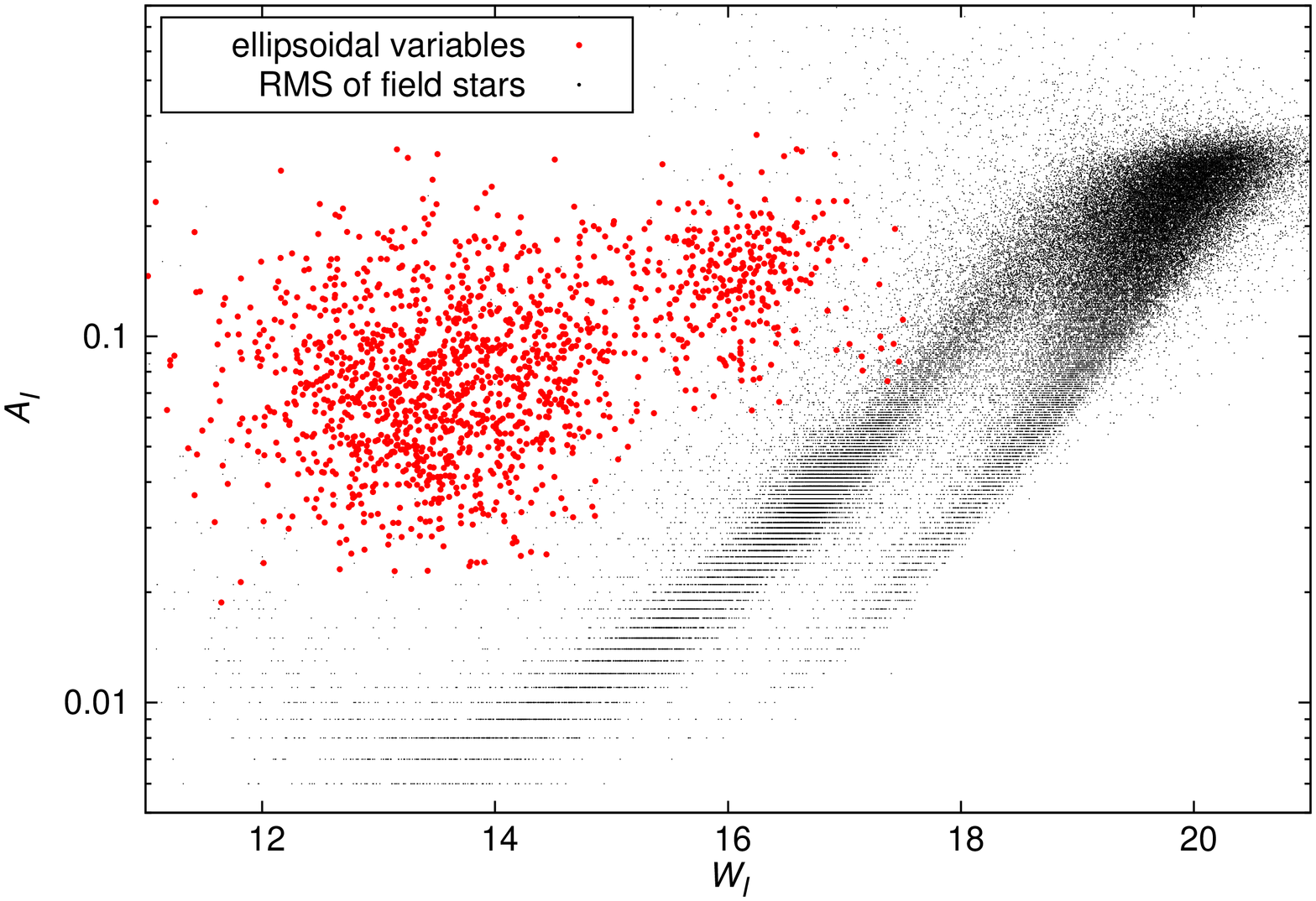}}
\FigCap{$I$-band amplitude of ellipsoidal binaries from our sample and RMS of field stars as a function of $W_I$.
The completeness of the sample may be lower for faint, low-amplitude stars.}
\end{figure}

\begin{figure}[htb]
\centerline{\includegraphics[angle=270,width=130mm]{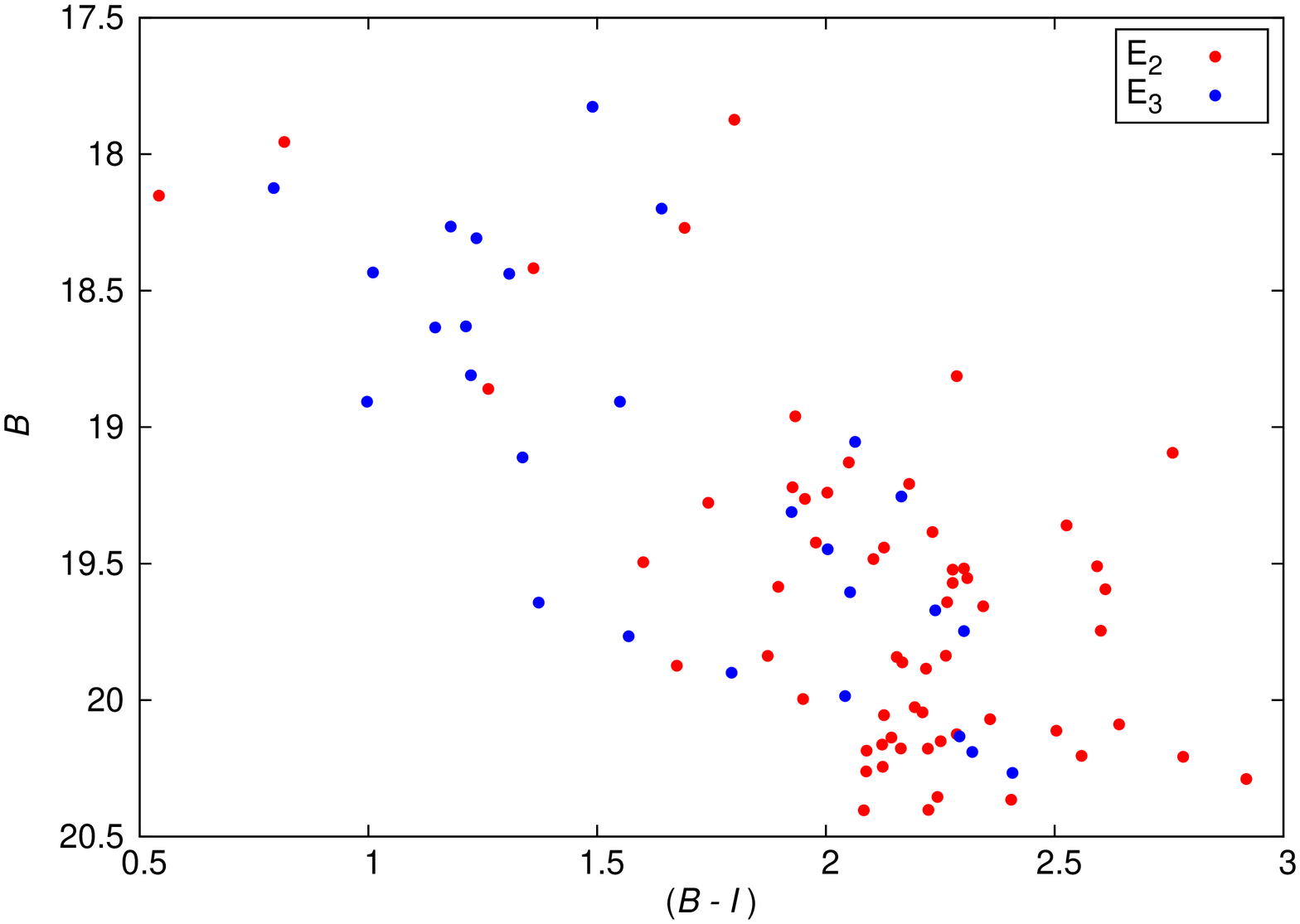}}
\FigCap{Color-magnitude diagram for E$_2$ (red points) and E$_3$ (blue points) subsequences. E$_3$ systems are significantly bluer than E$_2$.}
\end{figure}

Further difference between E$_2$ and E$_3$ can be found using $B$-band photometry obtained from 
the OGLE-II LMC photometric maps (Udalski \etal 2000) 
for some objects. Fig.~9 shows a $B$ \vs $(B-I)$ diagram for E$_2$ and E$_3$ systems only. Binaries in subsequence E$_3$ 
are significantly bluer and on average brighter in the $B$-band.

As an explanation to this effect, we propose the different nature of the non-ellipsoidal companion to the deformed red giant. 
It can be either another red giant (giant-giant system) or a smaller and bluer object, most likely a dwarf star close to turnoff or a subgiant (giant-dwarf system). In the first case the system will be very luminous $I$ and faint in $B$. This situation is observed for stars in E$_2$. 
The second case will be observed as fainter in $I$ but significantly brighter 
in $B$, as it is in subsequence E$_3$.

\begin{figure}[p]
\includegraphics[angle=270,width=62mm]{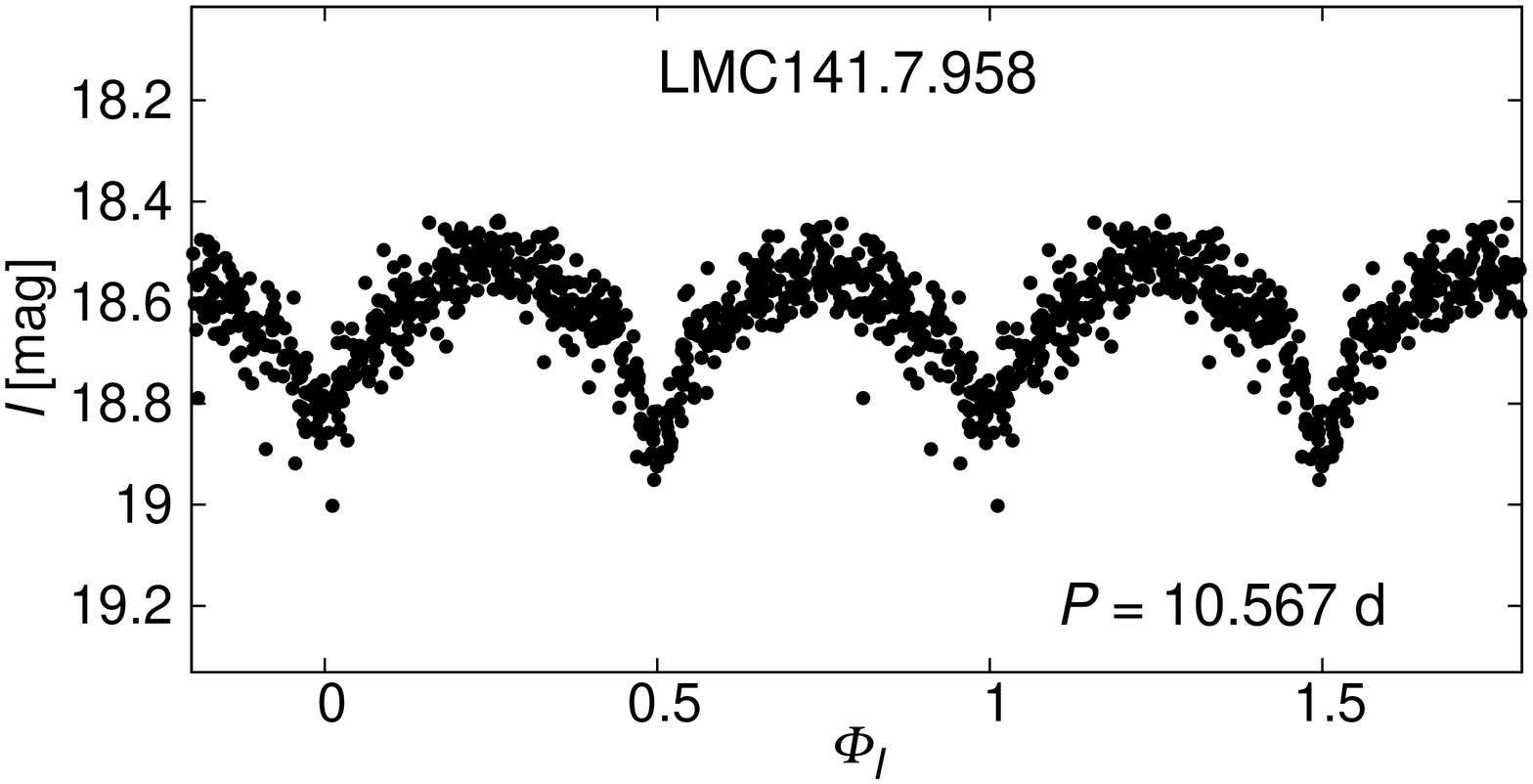}\hfill \includegraphics[angle=270,width=62mm]{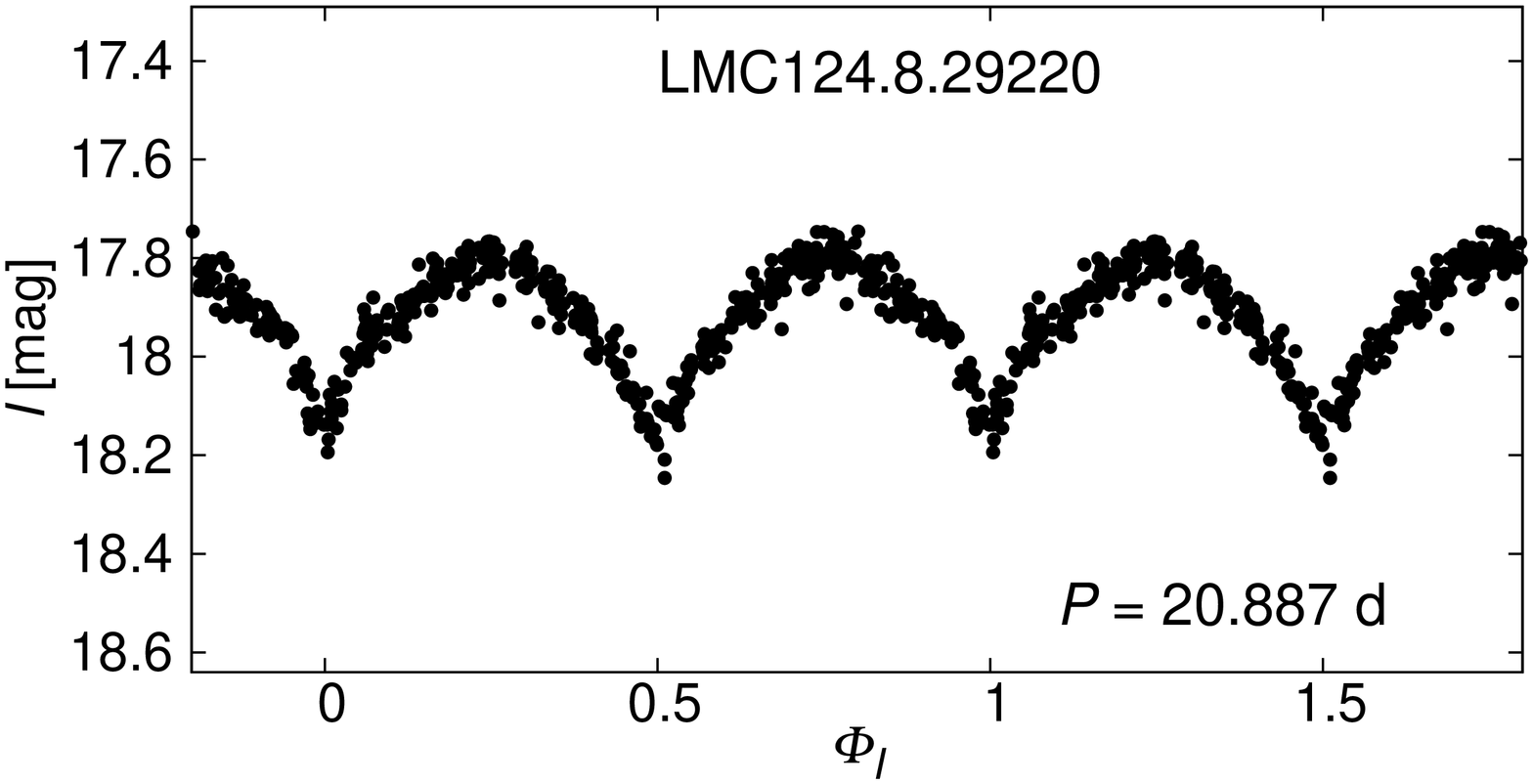} \\
\includegraphics[angle=270,width=62mm]{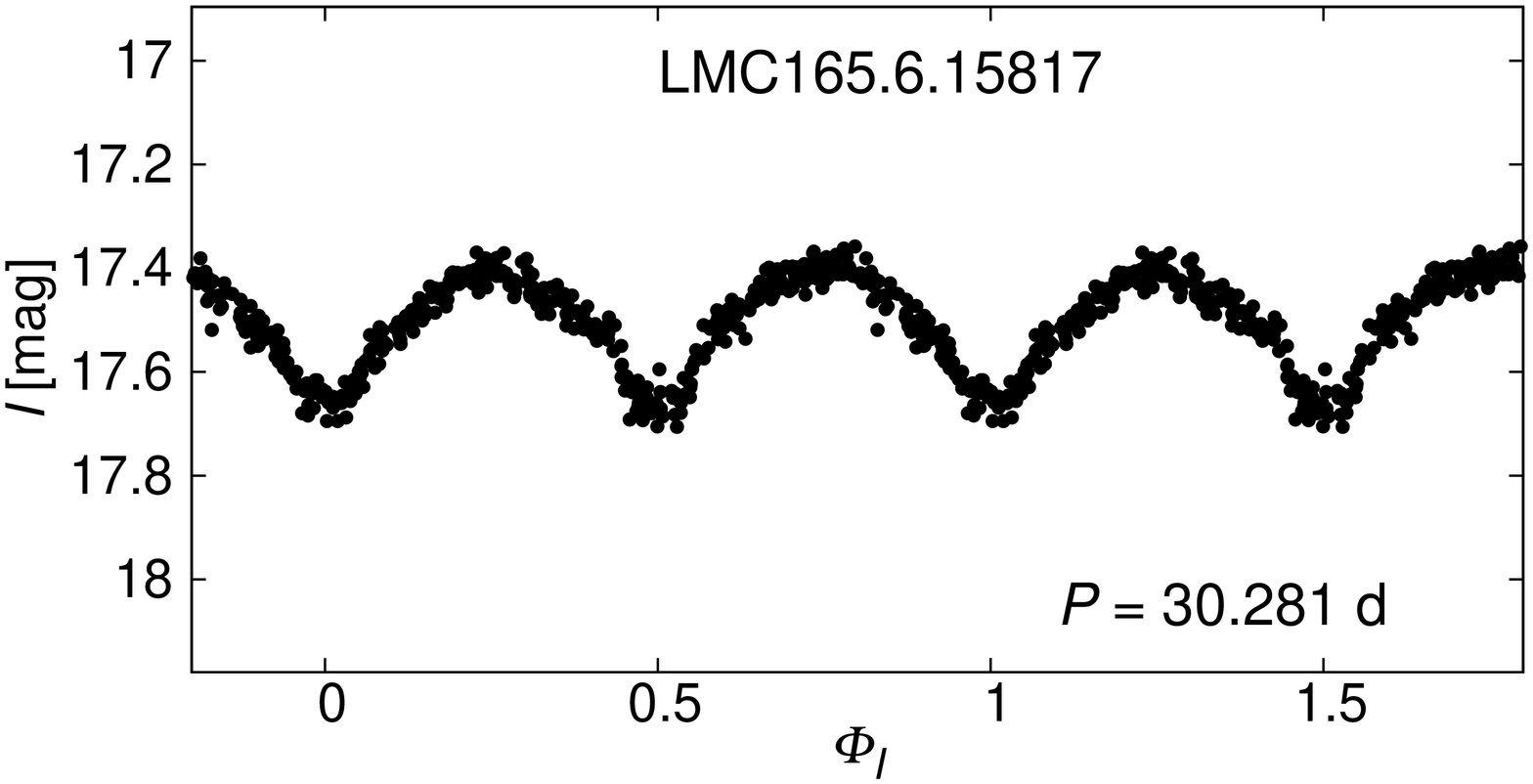}\hfill \includegraphics[angle=270,width=62mm]{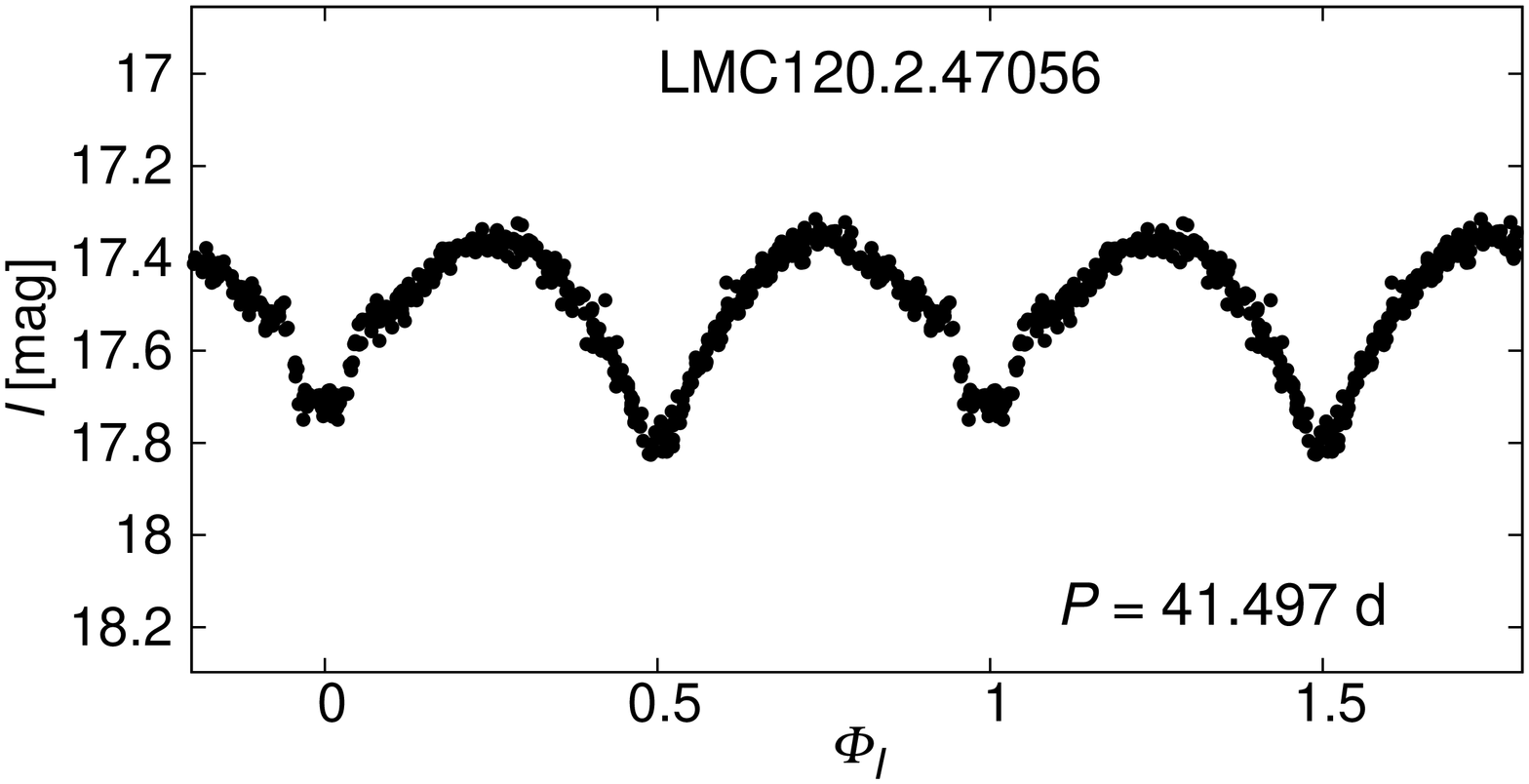}
\FigCap{Example light curves of eclipsing binaries from subsequence E$_2$. Both eclipses are of similar depth. These are likely giant-giant systems.}
\end{figure}

\begin{figure}[p]
\includegraphics[angle=270,width=62mm]{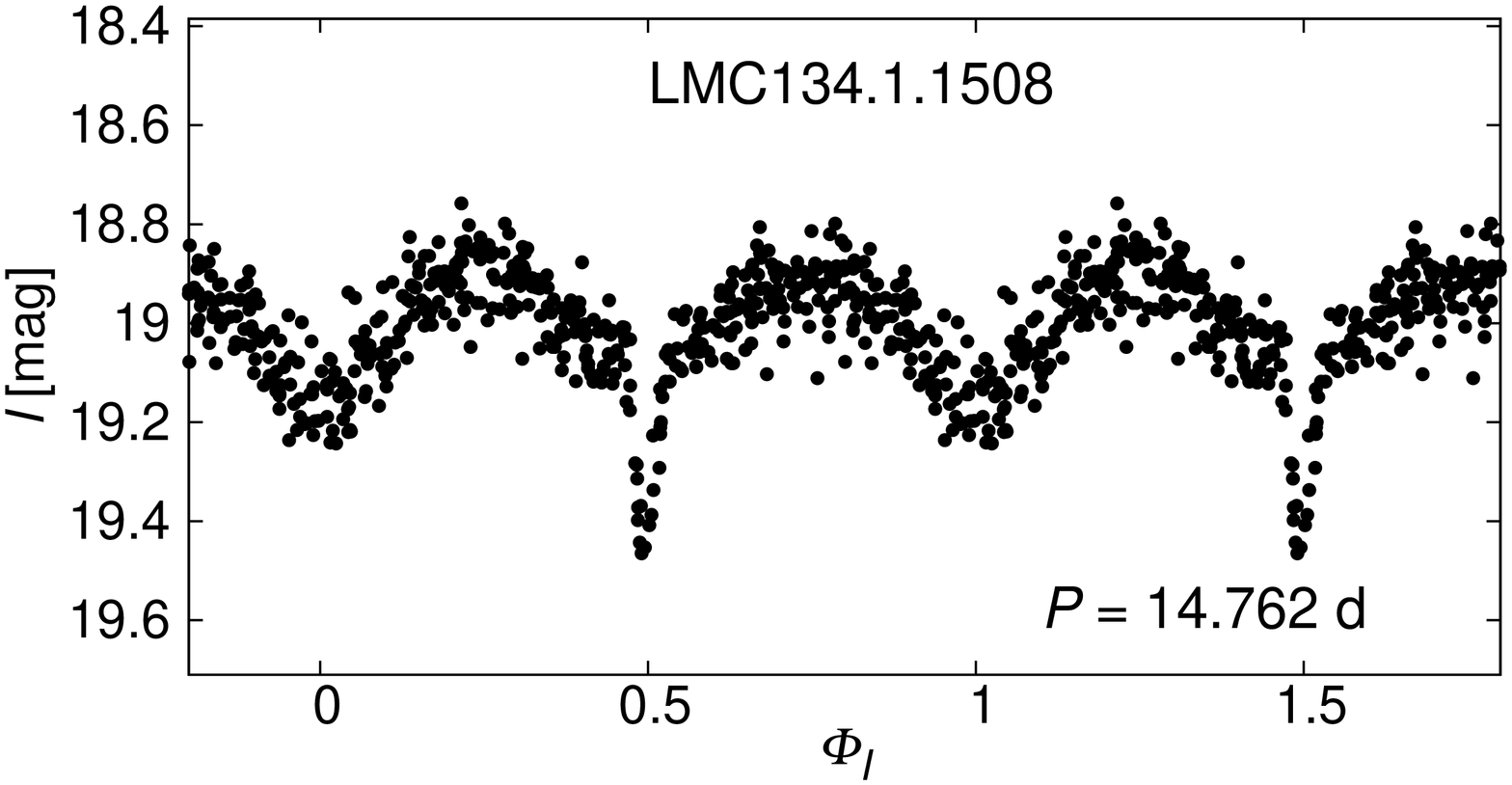}\hfill \includegraphics[angle=270,width=62mm]{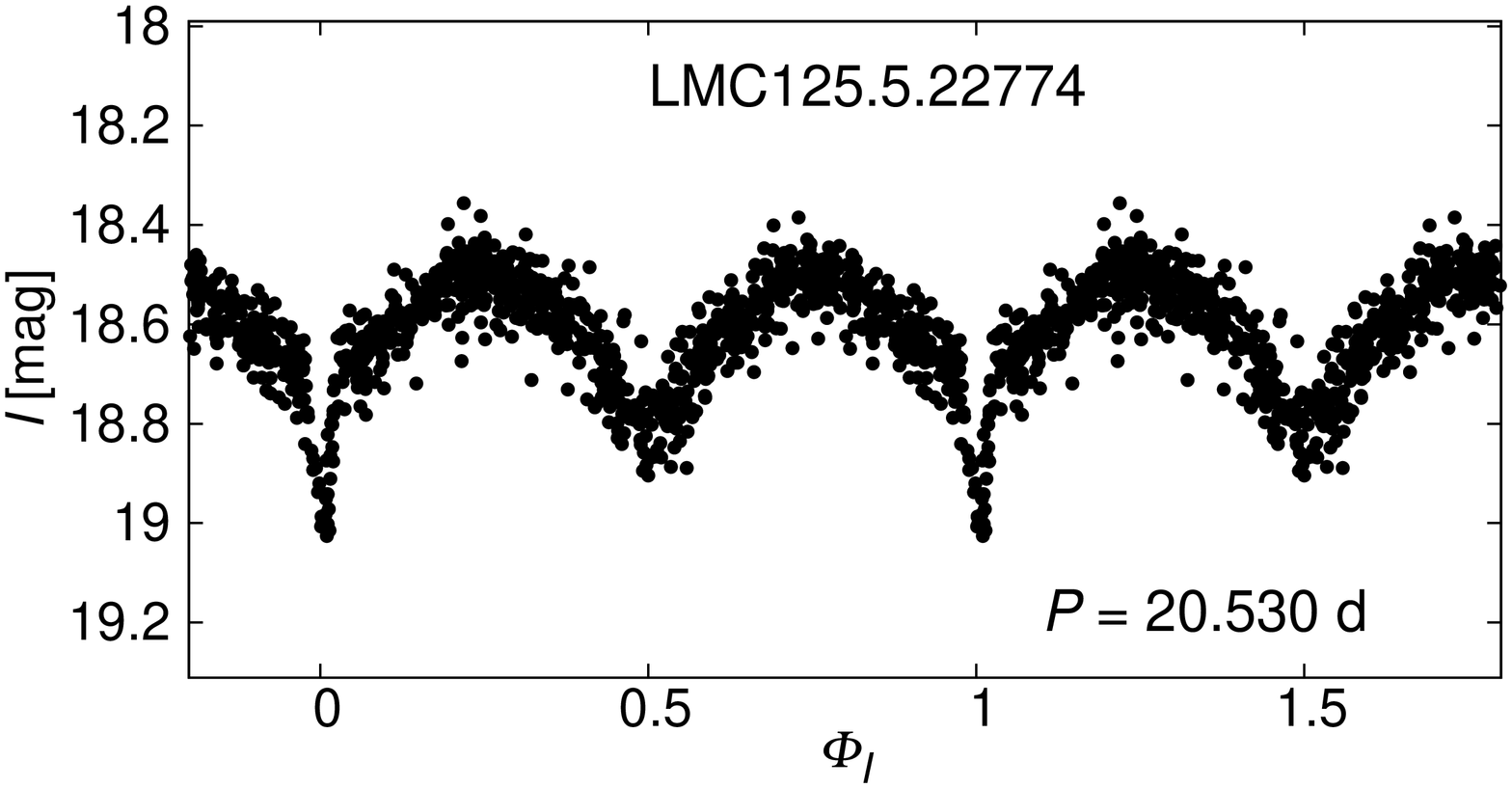} \\
\includegraphics[angle=270,width=62mm]{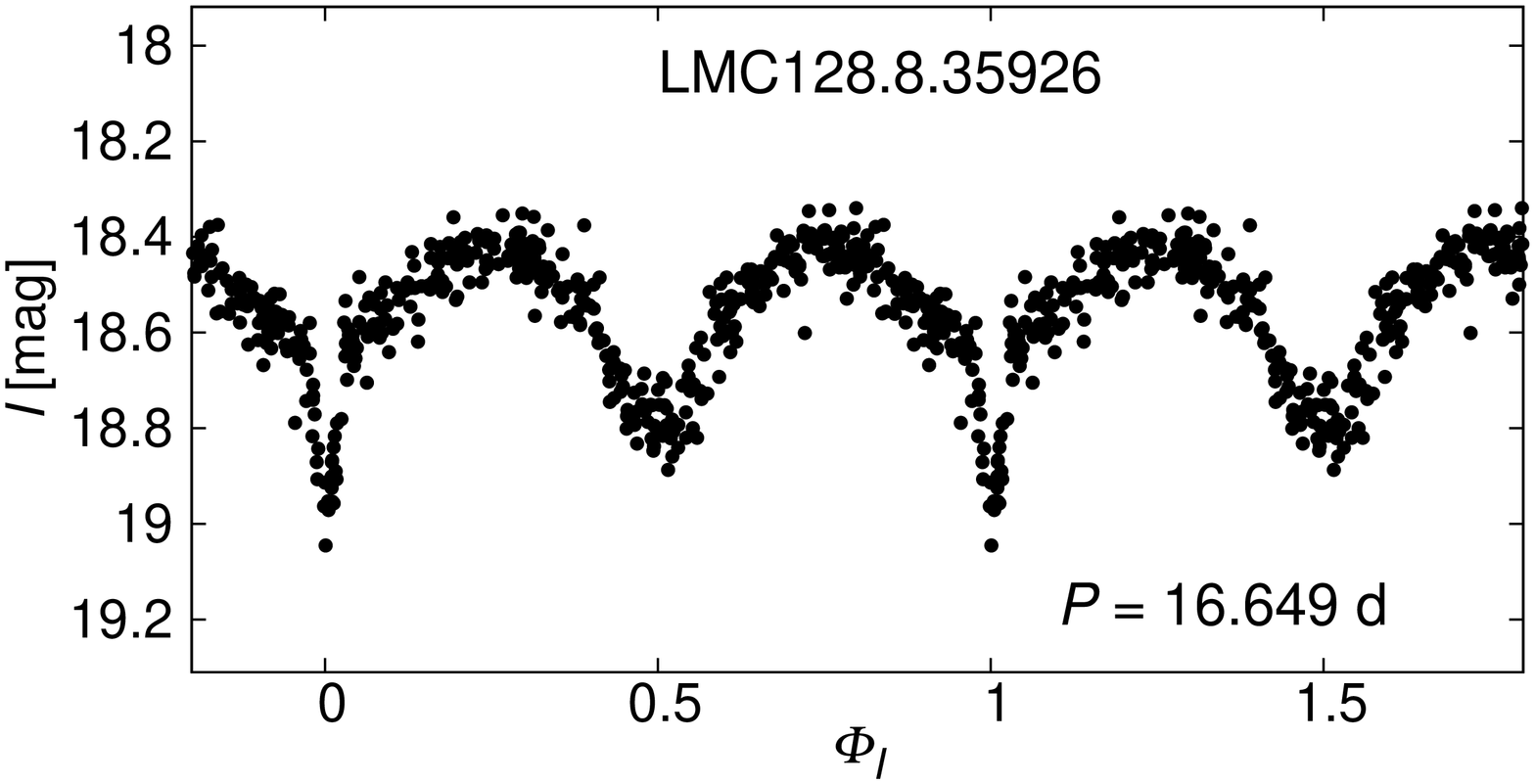}\hfill \includegraphics[angle=270,width=62mm]{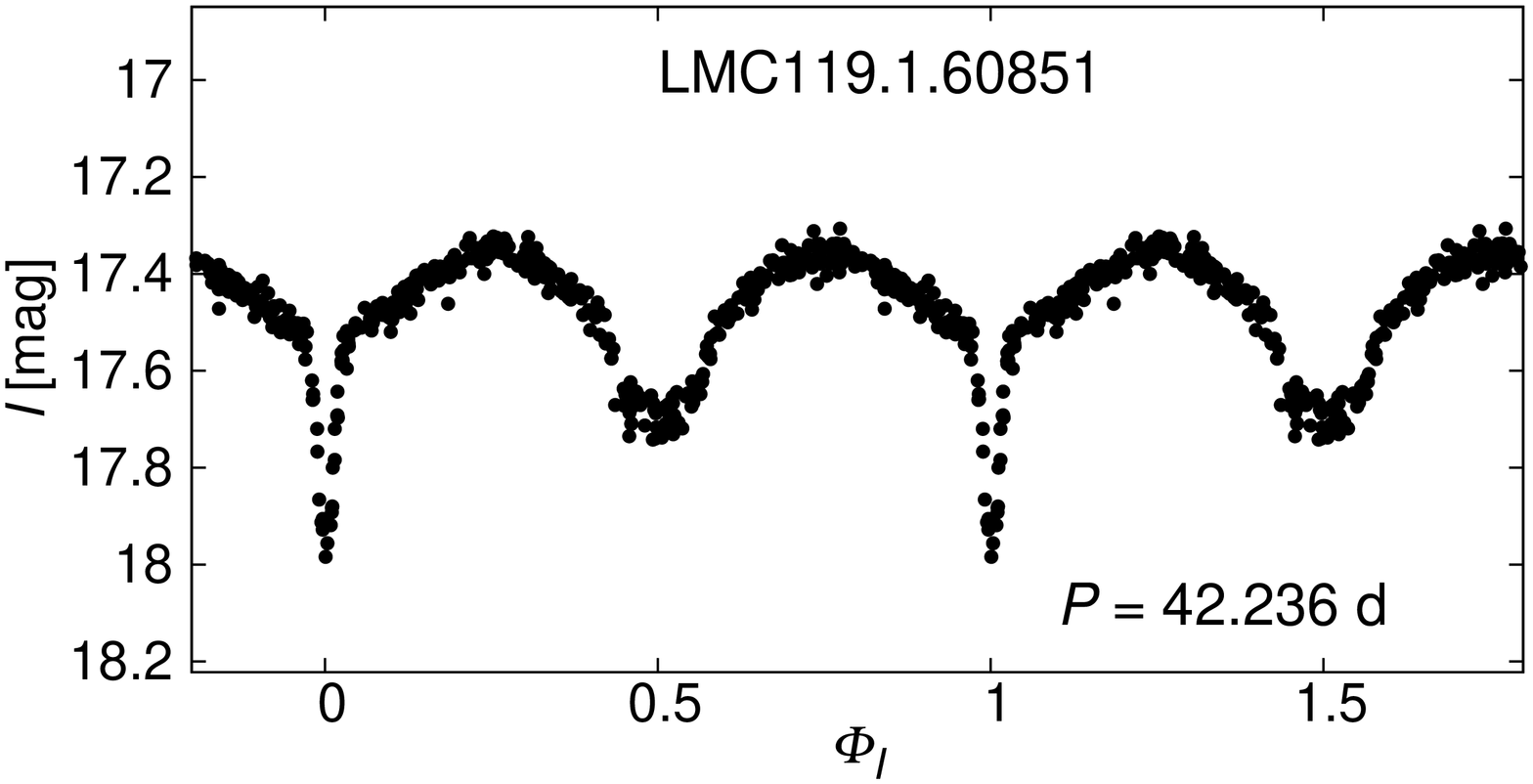}
\FigCap{Example light curves of eclipsing binaries from subsequence E$_3$. The primary eclipses are significantly deeper than the secondary ones. These are likely giant-dwarf systems.}
\end{figure}

\begin{figure}[htb]
\centerline{\includegraphics[angle=270,width=130mm]{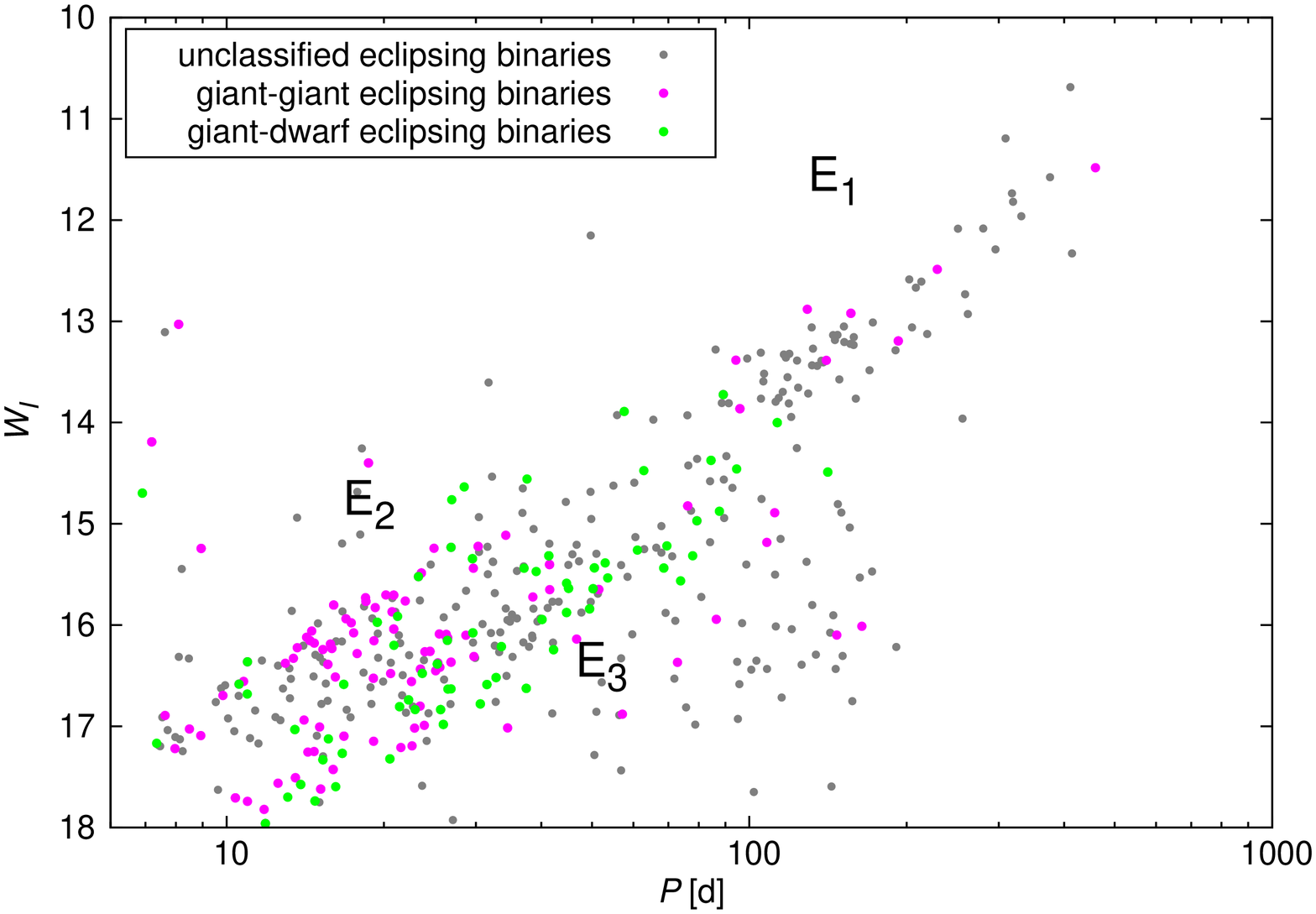}}
\FigCap{Position of likely giant-giant and giant-dwarf eclipsing binaries on the period-luminosity diagram. Giant-giant systems 
group mostly at E$_2$, while giant-dwarf at E$_3$.}
\end{figure}

Presented explanation is also consistent with the observed $I$-band amplitudes. 
The system in which most of the flux in $I$ comes from the ellipsoidal component
should have higher $I$-band amplitude than 
the system with two red components with only one showing ellipsoidal variability. 
This happens in subsequences E$_3$ and E$_2$, respectively.

While the light curves of non-eclipsing ellipsoidal binaries do not show significant differences between E$_2$ and E$_3$, 
the situation is different for eclipsing binaries (Graczyk \etal 2011).
Typical eclipsing light curves from E$_2$ and E$_3$ are shown in Fig.~10 and Fig.~11 respectively. 
Light curves with the primary eclipses significantly deeper than the secondary ones
are more common among stars located in the E$_3$ region.
This suggests that these systems have two components with different surface temperature. This is in agreement with our explanation.
The distribution of likely giant-giant and giant-dwarf systems is shown in Fig.~12. Once again, the separation between E$_2$ and E$_3$, 
composed of giant-giant and giant-dwarf binaries respectively, is clearly visible.

As it can be seen in Fig.~7, stars from subsequence E$_1$ with higher amplitudes occupy slightly different region than the ones with lower amplitudes.
High-amplitude stars are located in the upper part of this subsequence and seem to form a steeper relation.
On the other hand, stars from E$_3$ are fainter in $I$ and have on average higher amplitudes than stars from E$_2$.
Thus, we suspect that there are two types of ellipsoidal binaries. They are mixed at 
subsequence E$_1$ and separate into E$_2$ and E$_3$ in the region of shorter periods.
As suggested before, these two types corresponds to the giant-giant and giant-dwarf systems. 

With this assumption, we select high- and low-amplitude stars from the two regions.
The thresholds are $A_I > 0.15$~mag and $A_I < 0.05$~mag for systems with periods longer than 40~d,
and $A_I > 0.20$~mag and $A_I < 0.10$~mag for $P < 40$~d. The necessity of setting different thresholds results from the fact
that the systems with longer periods have statistically lower amplitudes.

We fit (rejecting outliers) a period-luminosity relations given
by Eq.~2 to the high- and low-amplitude stars separately: 

\begin{equation}
W_{I,{\rm high}}= -4.27 (\pm 0.12)\, {\log}P + 22.52 (\pm 0.24)
\end{equation}
\begin{equation}
W_{I,{\rm low}} = -2.78 (\pm 0.08)\, {\log}P + 19.30 (\pm 0.18) 
\end{equation}

The result is shown in Fig.~13.
The relation for high-amplitude (giant-dwarf) systems is significantly steeper than for the low-amplitude (giant-giant) ones.

For the short-period, faint objects the separation of the two relations is clear. However, for brighter systems the two relations cross and the objects 
forming them are much harder to separate. It is also possible that for the systems with a very bright ellipsoidal component, the influence of
the companion becomes negligible and the two relation merge into one.

\begin{figure}[htb]
\centerline{\includegraphics[angle=270,width=130mm]{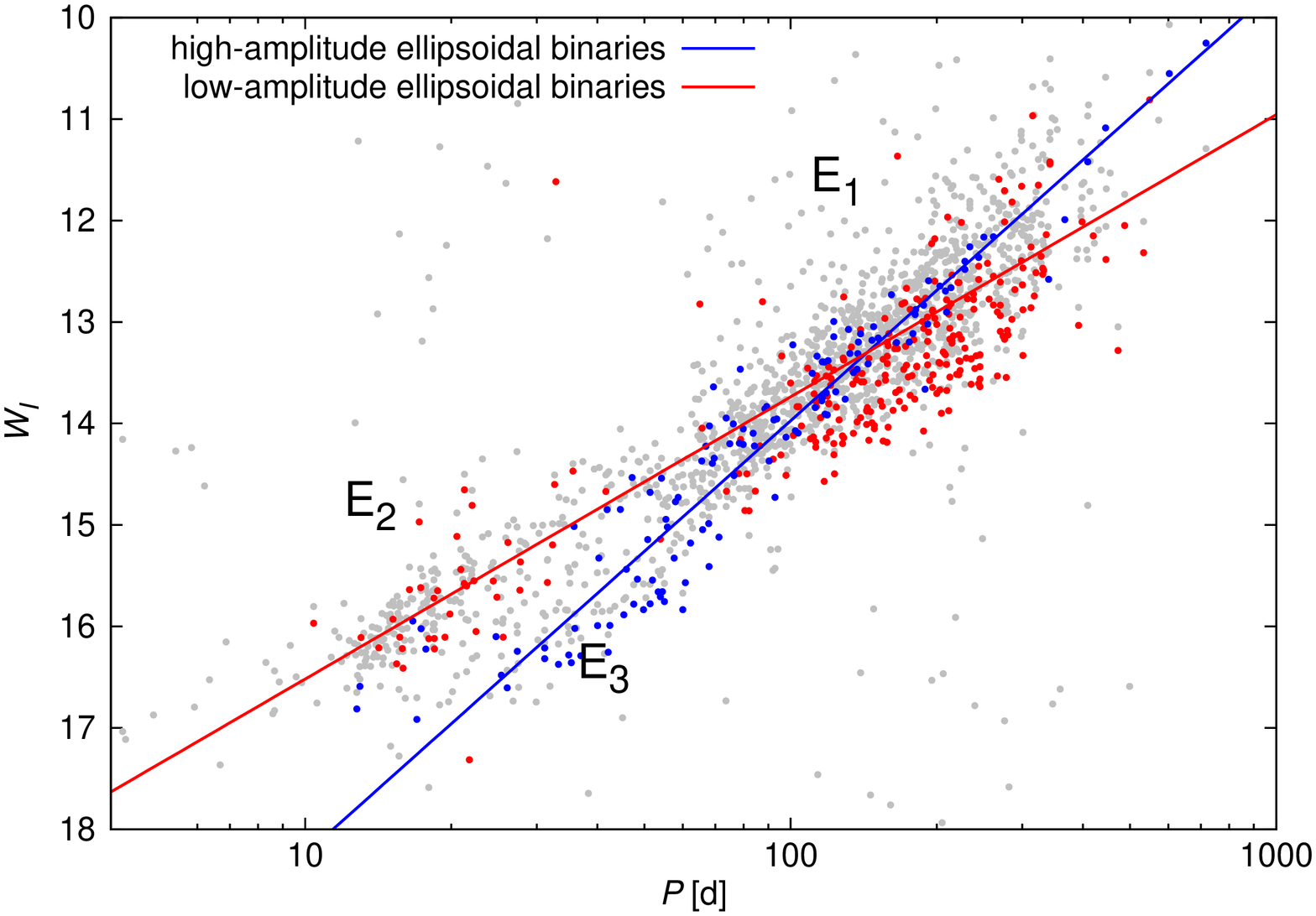}}
\FigCap{Period-luminosity relation for low- and high-amplitude ellipsoidal systems. These two groups are likely giant-giant and giant-dwarf binaries respectively.}
\end{figure}

\section{Ellipsoidal Binaries with Eccentric Orbit}

Another interesting type of ellipsoidal binaries are systems with eccentric orbit. 
Large sample of such systems discovered during the OGLE-II phase was presented by Soszy{\'n}ski \etal (2004).
Our OGLE-III sample contains 271 such objects. The shape of the light curve varies depending on the eccentricity and orbit configuration relative to the line-of-sight. It is close to sinusoidal, with one
minimum broader than another for systems with slightly eccentric orbit. For systems with increasing eccentricity, the light curve becomes heavily asymmetric. 
Example light curves of ellipsoidal variables with eccentric orbits are shown in Fig.~14.

\begin{figure}[p]
\includegraphics[angle=270,width=62mm]{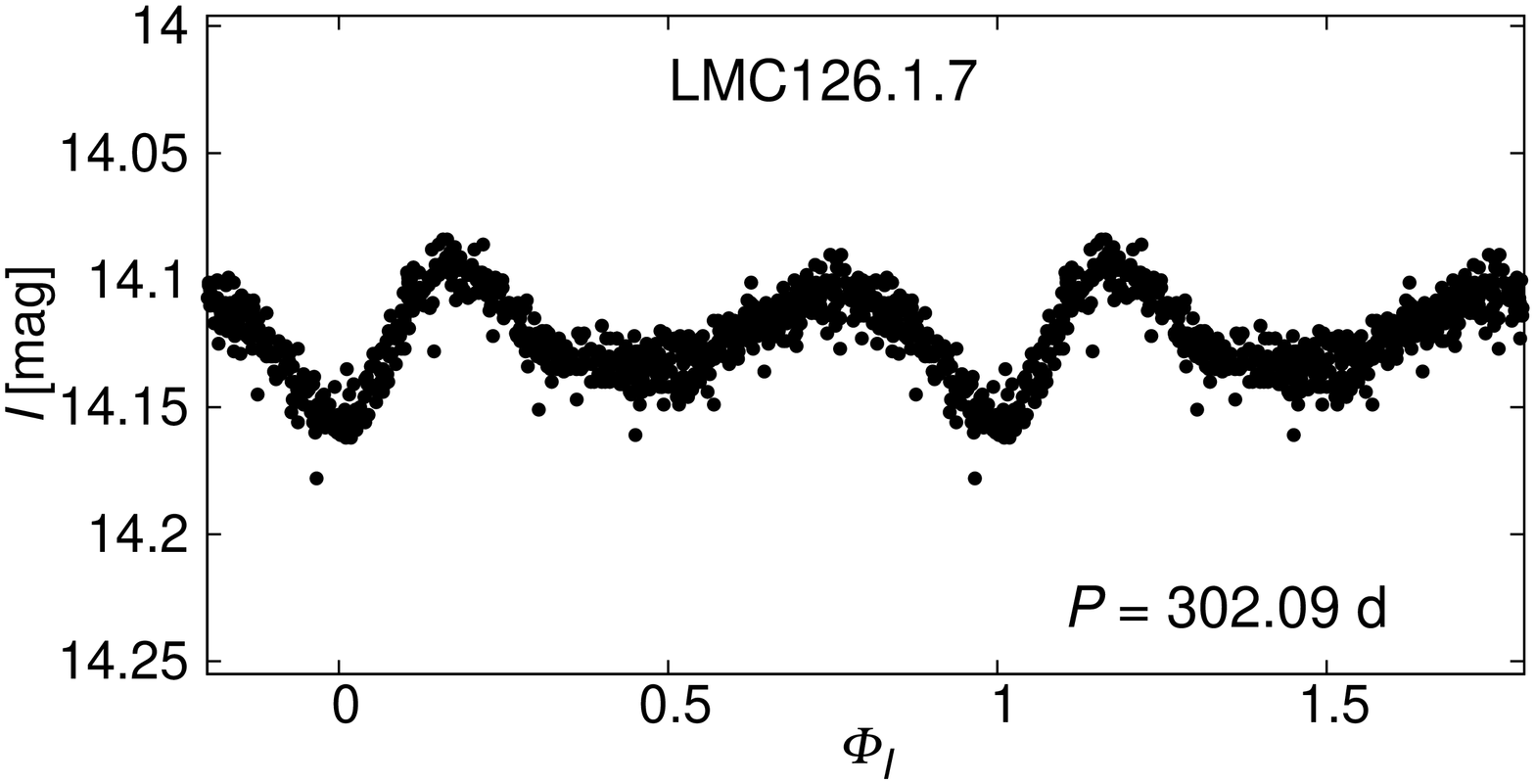}\hfill \includegraphics[angle=270,width=62mm]{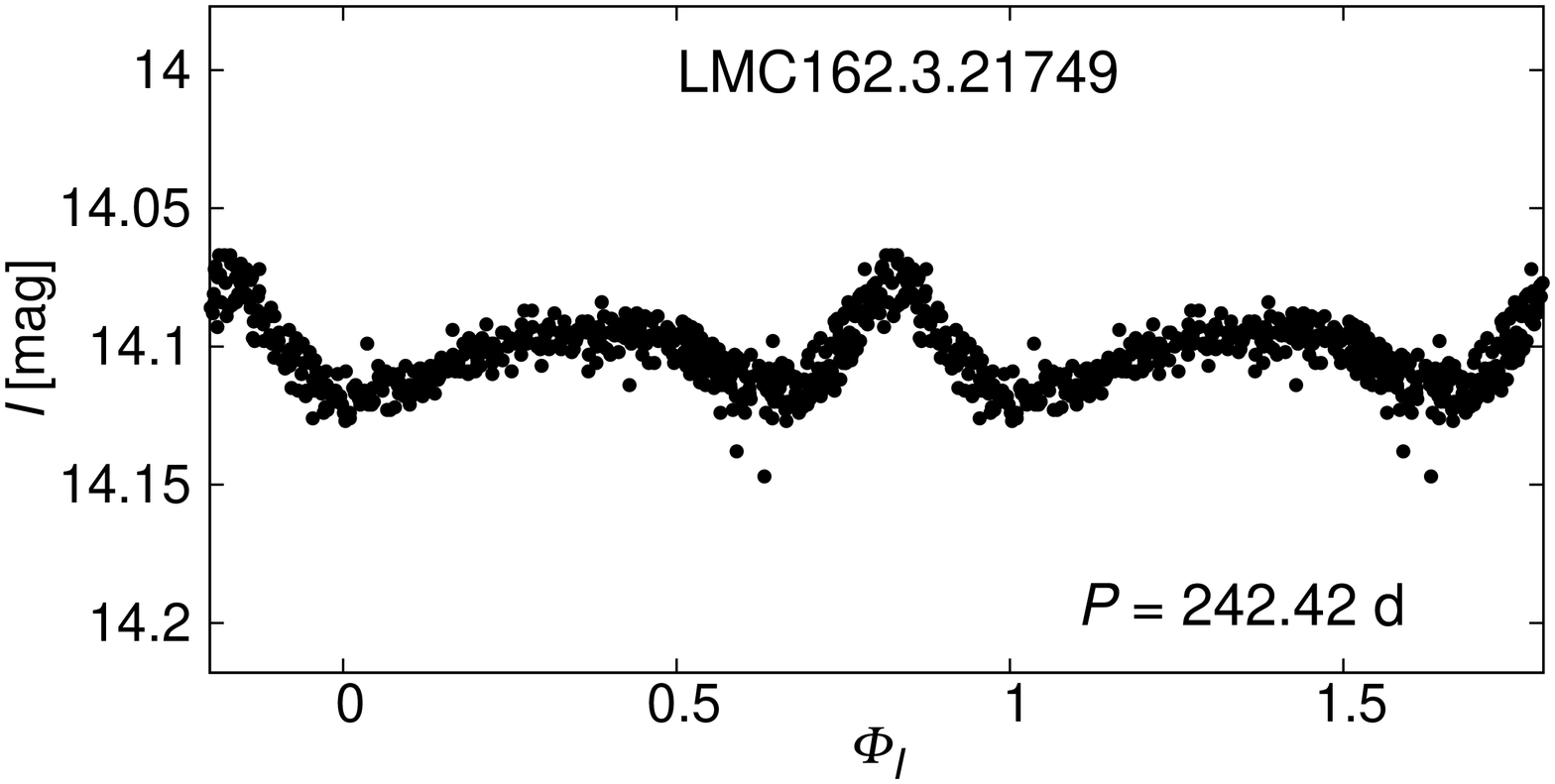} \\
\includegraphics[angle=270,width=62mm]{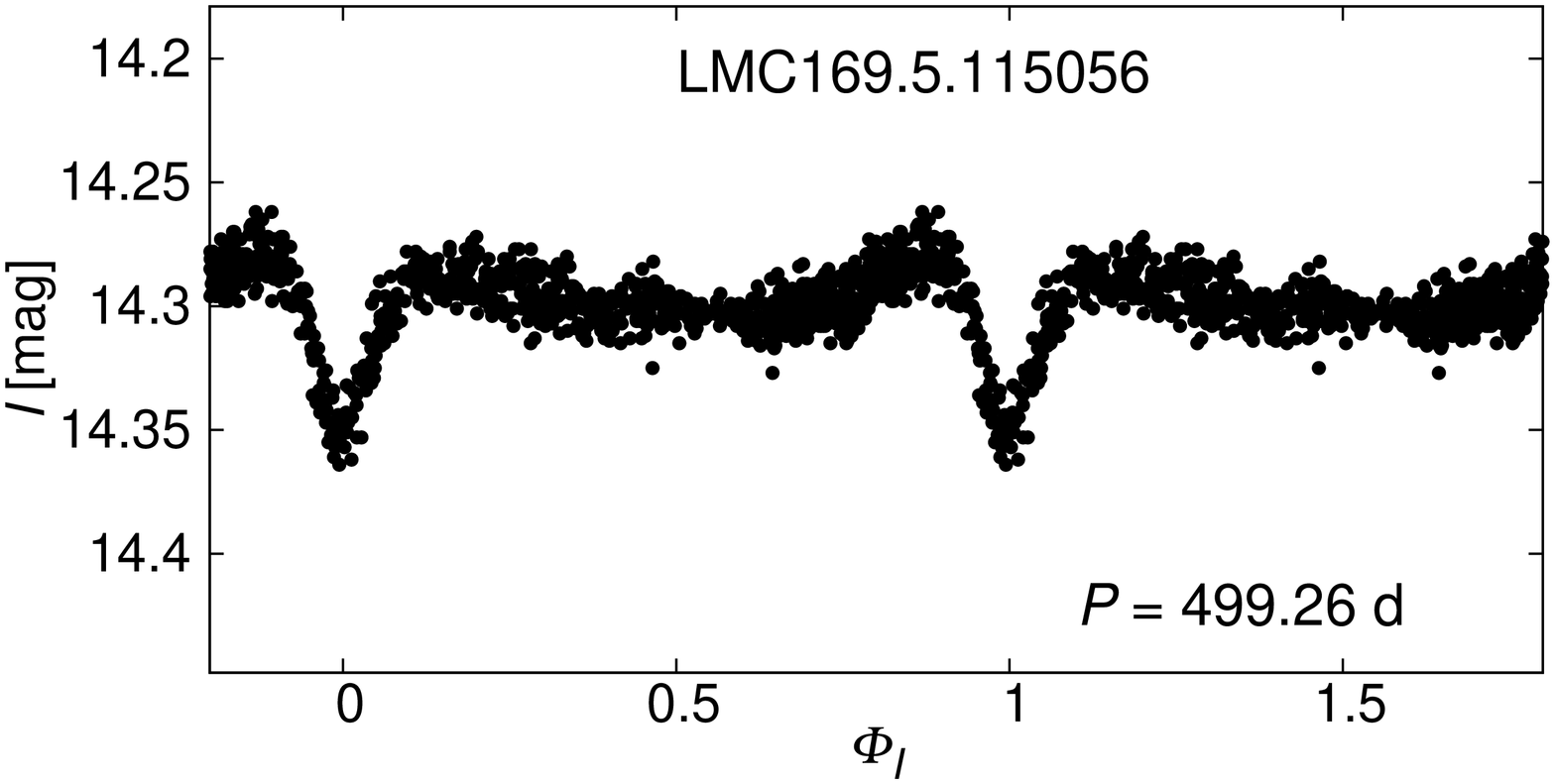}\hfill \includegraphics[angle=270,width=62mm]{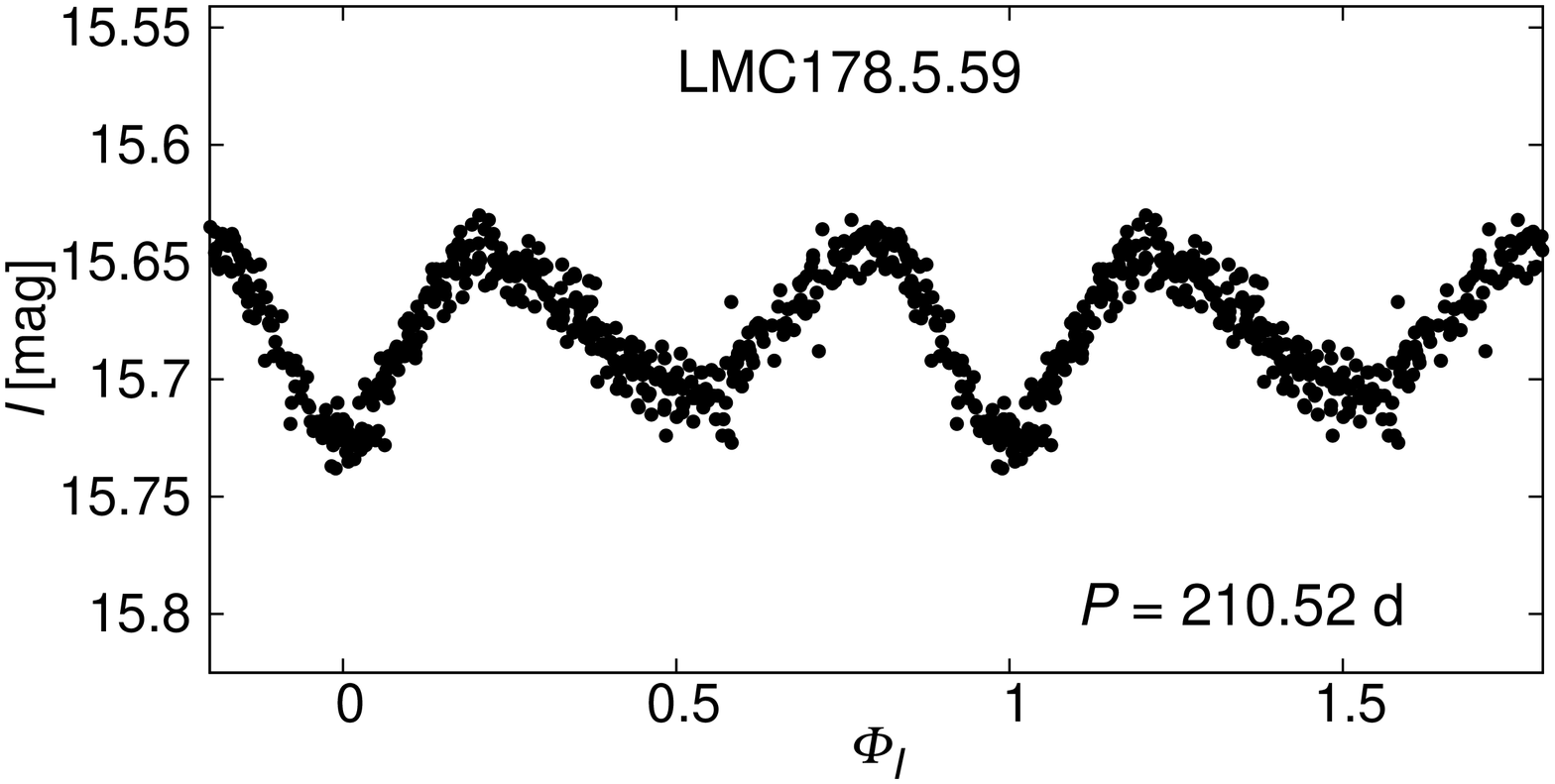}  
\FigCap{Example light curves of ellipsoidal systems with eccentric orbit.}
\end{figure}

\begin{figure}[htb]
\centerline{\includegraphics[angle=270,width=130mm]{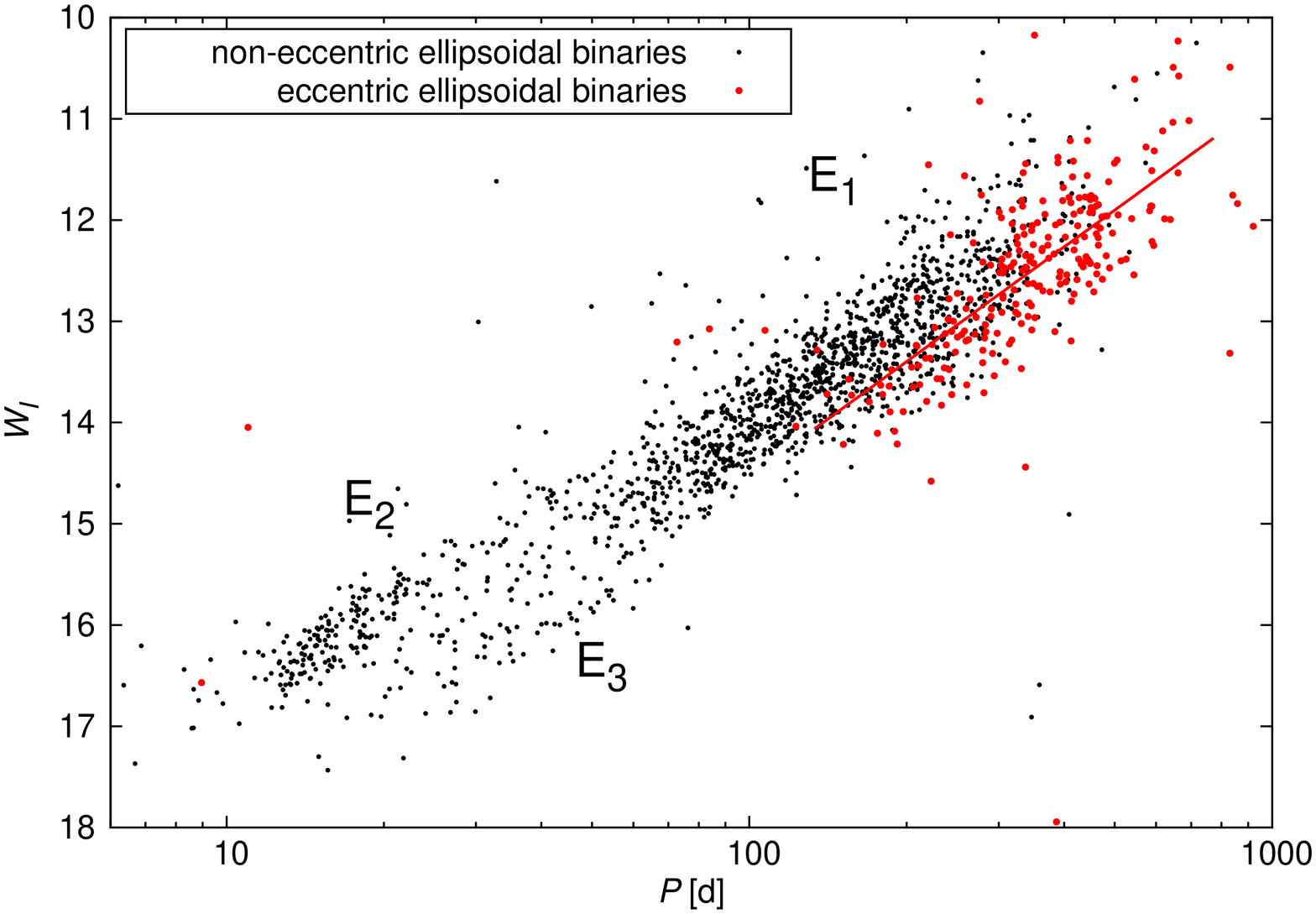}}
\FigCap{Position of the systems with eccentric orbit on the ${\log}P-W_I$ diagram.}
\end{figure}

Systems with eccentric orbits have long orbital periods, in most cases $P > 150$~d. They lie to the right from sequence E$_1$, 
as shown in Fig.~15. We obtain the following linear period-luminosity relation for eccentric systems:

\begin{equation}
W_{I,{\rm eccen}} = -3.76 (\pm 0.22)\, {\log}P + 22.05 (\pm 0.55)
\end{equation}

\section{Summary and Conclusions}    

We independently searched for ellipsoidal binary stars in the OGLE-III LMC data.
Our sample consists of 5334 light curves, in the region of sequence E, consistent with ellipsoidal variability, out of which we 
identify 1565 high-confidence objects with two different-depth minima in the phased light curves.

We find that ellipsoidal binary systems clearly group around two ranges of orbital periods, one 
with a maximum around 12~d and the other one around 130~d. The first maximum corresponds
to RC stars, while the second to RGB stars. 
We show that stars from previously known sequence E in the period-luminosity diagram,
in fact separate into three groups, 
labeled as E$_1$, E$_2$, and E$_3$. Subsequence E$_1$ roughly corresponds to systems with period $P > 40$~d, while
E$_2$ and E$_3$ to systems with $P < 40$~d.
All three subsequences are also visible for close binary systems with eclipses. 

We suggest that subsequence E$_3$ is formed of giant-dwarf binaries,
while E$_2$ of giant-giant systems. Analysis of $B$-band magnitudes and shapes of
eclipsing light curves support this claim.

Systems from subsequence E$_1$ with low and high amplitudes form different period-luminosity 
relations. High-amplitude stars lie in the upper
region of this subsequence and form a steeper relation.
In the short-period regime ($P<40$~d) low-amplitude stars mostly lie on subsequence E$_2$, while 
high-amplitude stars rather concentrate around subsequence E$_3$ located below E$_2$.
We suspect that there are two types of ellipsoidal binaries 
which merge in the subsequence E$_1$ and separate into E$_2$ and E$_3$ for shorter periods.
For high- and low-amplitude stars we derive two separate relations.
The relation for high-amplitude objects is significantly steeper. 

From the whole sample of 5334 ellipsoidal stars, 271 ellipsoidal binaries have eccentric orbits. Those stars 
have on average longer periods than non-eccentric systems of the same brightness.

\section*{Acknowledgments}
We would like to thank prof. K.~St\k{e}pie{\'n} for fruitful discussion and invaluable comments.
We thank Z.~Ko{\l}aczkowski for providing {\sc Fnpeaks} 
period search code. 

This work has been supported by the Polish National Science Center grant No.
DEC-2011/03/B/ST9/02573.
The OGLE project has received funding from the European Research Council
under the European Community's Seventh Framework Programme
(FP7/2007-2013) / ERC grant agreement no. 246678 to AU.

\end{document}